\documentclass[twocolumn]{aastex631}
\usepackage{amsmath}
\usepackage{multirow}
\usepackage{url}
\usepackage{color}
\usepackage{hyperref}
\usepackage{fnbreak}
\usepackage{subfigure}
\usepackage{acronym}
\usepackage{xspace}
\newcommand{\sref}[1]{Sec.~\ref{#1}}
\newcommand{\aref}[1]{Appendix~\ref{#1}}
\newcommand{\fref}[1]{Fig.~\ref{#1}}
\newcommand{\tref}[1]{Table~\ref{#1}}

\newcommand{\eref}[1]{equation~\eqref{#1}}
\newcommand{\Sref}[1]{Section~\ref{#1}}

\newcommand{\Tref}[1]{Table~\ref{#1}}
\newcommand{\Eref}[1]{Equation~\eqref{#1}}

\newcommand{\mc}[1]{\mathcal{#1}}
\newcommand{\F}{\mc{F}}
\newcommand{\abs}[1]{\left\lvert#1\right\rvert}
\newcommand{\smallabs}[1]{\lvert#1\rvert}
\newcommand{\ev}[1]{E\left[#1\right]}
\newcommand{\levarm}{r}
\newcommand{\Rstar}{R_*}

\newcommand{\Tasc}{t_{\text{asc}}}
\newcommand{\TascThen}{\Tasc} \newcommand{\TascNow}{\Tasc'} \newcommand{\Tsft}{T_{\text{sft}}}
\newcommand{\Tmax}{T_{\text{max}}}

\newcommand{\Trun}{T_{\text{run}}}

\newcommand{\Tdrift}{T_{\text{drift}}}
\newcommand{\fdotdrift}{\smallabs{\dot{f}}_{\text{drift}}}
\newcommand{\Porb}{P_{\text{orb}}}
\newcommand{\PorbShear}{\tilde{P}}

\newcommand{\rhomax}{\rho_{\text{max}}}
\newcommand{\curlyrho}{\vartheta}

\newcommand{\un}[1]{\,\text{#1}}
\newcommand{\tbin}{\mathfrak{t}}
\newcommand{\code}[1]{\texttt{#1}\xspace}

\newcommand{\coord}{coordinate}

\newcommand{\naive}{na\"{\i}ve}
\newcommand{\dcc}{LIGO-P2500260-v11}

\newcommand{\LSCRaHr}{16}
\newcommand{\LSCRaMin}{19}
\newcommand{\LSCRaSec}{55.0850}
\newcommand{\LSCNegDecDeg}{15}
\newcommand{\LSCNegDecMin}{38}
\newcommand{\LSCNegDecSec}{24.9}
\newcommand{\KillPorbSec}{68023.919}
\newcommand{\KillPorbHr}{18.9}
\newcommand{\KilldPorbSec}{0.017}
\newcommand{\KillTascGPS}{1078153676}
\newcommand{\KillTascDayTime}{2014--Mar--06 15:07:40\,UTC}

\newcommand{\KillTascYear}{2014}
\newcommand{\KilldTascGPS}{33}
\newcommand{\OivaTascGPS}{1379295565}
\newcommand{\OivaTascDayTime}{2023--Sep--21 01:39:07\,UTC}

\newcommand{\OivaTascYear}{2023}
\newcommand{\OivadTascGPS}{82}
\newcommand{\OiiiTmaxMin}{240}

\newcommand{\OiTmaxMax}{25920}
\newcommand{\OiiTmax}{68400}
\newcommand{\OivaTmax}{86400}
\newcommand{\OivadPtildeSec}{0.006}
\newcommand{\OivadPorbdTascShear}{1.89\times 10^{-4}}
\newcommand{\KillAsiniMin}{1.44}
\newcommand{\KillAsiniMax}{3.25}
\newcommand{\htorque}{3.4\times 10^{-26}}
\newcommand{\OivaStartGPS}{1368975618}
\newcommand{\OivaStartDayTime}{2023--May--24 15:00:00\,UTC}

\newcommand{\OivaEndGPS}{1389456018}
\newcommand{\OivaEndDayTime}{2024--Jan--16 16:00:00\,UTC}

\newcommand{\OivbStartGPS}{1396969218}
\newcommand{\OivbStartDayTime}{2024--Apr--12 15:00:00\,UTC}

\newcommand{\OivbEndGPS}{1422118818}
\newcommand{\OivbEndDayTime}{2025--Jan--28 17:00:00\,UTC}

\newcommand{\OivciiEndGPS}{1447516818}
\newcommand{\OivciiEndDayTime}{2025--Nov--18 16:00:00\,UTC}

\newcommand{\OivaTrun}{2.05\times 10^{7}}
\newcommand{\OivaSWLoss}{0.251}
\newcommand{\EccLoss}{0.021}
\newcommand{\tpPct}{99.6}
\newcommand{\OivaNorb}{4427}
\newcommand{\onesigpct}{39.3}
\newcommand{\twosigpct}{86.5}
\newcommand{\threesigpct}{98.9}
\newcommand{\mismatchMax}{0.25}

\newcommand{\numUnvetoed}{51}
\newcommand{\numInjTotal}{350}

\newcommand{\numInjFollowup}{289}
\newcommand{\numInjUnvetoed}{277}
\newcommand{\numInjVetoedi}{8}
\newcommand{\numInjVetoedii}{2}
\newcommand{\numInjVetoediii}{2}
\newcommand{\outlierratio}{1.75}
\newcommand{\outlierf}{129.20}
\newcommand{\outlierrho}{7.84}
\newcommand{\outlierrhoo}{5.91}
\newcommand{\outlierrhoi}{5.93}
\newcommand{\outlierrhoii}{9.58}
\newcommand{\outlierrhoiii}{11.70}
\newcommand{\outlierratiotwoone}{1.61}
\newcommand{\outlierratiotretwo}{1.22}
\newcommand{\injminrhoiii}{15.96}
\newcommand{\fmina}{25}
\newcommand{\fmaxa}{50}
\newcommand{\rhothresha}{7.2}
\newcommand{\fminb}{50}
\newcommand{\fmaxb}{100}
\newcommand{\rhothreshb}{7.3}
\newcommand{\fminc}{100}
\newcommand{\fmaxc}{150}
\newcommand{\rhothreshc}{7.5}
\newcommand{\fmind}{150}
\newcommand{\fmaxd}{200}
\newcommand{\rhothreshd}{7.6}
\begin{document}
\reportnum{\dcc}
\title{Sub-Torque-Balance Upper Limits on Continuous Gravitational Waves from Scorpius X-1}
\suppressAffiliations
\author[0000-0003-4786-2698]{A.~G.~Abac}
\affiliation{Max Planck Institute for Gravitational Physics (Albert Einstein Institute), D-14476 Potsdam, Germany}
\author{I.~Abouelfettouh}
\affiliation{LIGO Hanford Observatory, Richland, WA 99352, USA}
\author{F.~Acernese}
\affiliation{Dipartimento di Farmacia, Universit\`a di Salerno, I-84084 Fisciano, Salerno, Italy}
\affiliation{INFN, Sezione di Napoli, I-80126 Napoli, Italy}
\author[0000-0002-8648-0767]{K.~Ackley}
\affiliation{University of Warwick, Coventry CV4 7AL, United Kingdom}
\author{A.~Adam}
\affiliation{OzGrav, University of Western Australia, Crawley, Western Australia 6009, Australia}
\author[0000-0001-5525-6255]{C.~Adamcewicz}
\affiliation{OzGrav, School of Physics \& Astronomy, Monash University, Clayton 3800, Victoria, Australia}
\author[0009-0004-2101-5428]{S.~Adhicary}
\affiliation{The Pennsylvania State University, University Park, PA 16802, USA}
\author{D.~Adhikari}
\affiliation{Max Planck Institute for Gravitational Physics (Albert Einstein Institute), D-30167 Hannover, Germany}
\affiliation{Leibniz Universit\"{a}t Hannover, D-30167 Hannover, Germany}
\author[0000-0002-4559-8427]{N.~Adhikari}
\affiliation{University of Wisconsin-Milwaukee, Milwaukee, WI 53201, USA}
\author[0000-0002-5731-5076]{R.~X.~Adhikari}
\affiliation{LIGO Laboratory, California Institute of Technology, Pasadena, CA 91125, USA}
% ...
\collaboration{1}{The LIGO Scientific Collaboration, the Virgo Collaboration, and the KAGRA Collaboration}
\author[0000-0002-0440-9597]{T.~L.~Killestein}
\affiliation{University of Warwick, Coventry CV4 7AL, United Kingdom}
\author[0000-0003-0771-4746]{D.~Steeghs}
\affiliation{University of Warwick, Coventry CV4 7AL, United Kingdom}
\author[0000-0001-5031-0128]{J.~Casares}
\affiliation{Instituto de Astrof\'{\i}sica de Canarias, E-38205 La Laguna, Tenerife, Spain}
\affiliation{Departamento de Astrof\'{\i}sica, Universidad de La Laguna, E-38206 La Laguna, Tenerife, Spain}
\author{D.~K.~Galloway}
\affiliation{OzGrav, School of Physics \& Astronomy, Monash University, Clayton 3800, Victoria, Australia}
\collaboration{1}{Precision Ephemerides for Gravitational-Wave Searches (PEGS) Project}
 \date{2026 July 7}
\begin{abstract}
  We present the results of a search for continuous
  gravitational waves from the low-mass X-ray binary Scorpius X-1
  using LIGO data from the first part of the fourth LIGO-Virgo-KAGRA
  observing run.  By applying the
  resampling version of the cross-correlation pipeline to search for
  signal frequencies $f_0$ between $25$ and $200\un{Hz}$
  (corresponding to neutron star spin frequencies of $12.5$ to
  $100\un{Hz}$ for GW due to triaxiality, or {$\sim15-20$}
  to {$\sim120-150\un{Hz}$} for GW due to $r$-modes),
we
set upper limits below the standard torque balance
  level, independent of neutron star spin inclination, for
  $50\un{Hz}\lesssim f_0\lesssim200\un{Hz}$.
While uncertainties in the modelling of torque and equation of state
  limit the strength of our inference, our results nonetheless argue
  against torque balance in this spin range for a neutron star
  described by a hadronic equation of state.
The most sensitive upper limits on the gravitational wave amplitude
  $h_0$, at the upper end of the frequency band searched, approach
  {$5\times10^{-26}$} marginalized over inclination angle and
  {$2\times10^{-26}$} assuming the most favorable
  inclination.  The marginalized upper limits correspond to a
  sensitivity depth of $70-75\un{Hz}^{-1/2}$, improving sensitivity
  considerably over previous searches.  Expressed as constraints on
  the triaxial deformation of the neutron star, the limits correspond
  to an ellipticity of {$3\times10^{-5}$} if the GW frequency
  $f_0$ is $75\un{Hz}$ and {$3\times10^{-6}$} if
  $f_0=200\un{Hz}$, approaching deformations which could be
  supported by ordinary nuclear matter.  Outliers from the search were
  ruled out as potential signals by a combination of hierarchical
  followup and analysis of additional data from later in the observing run.
\end{abstract}

\acrodef{NS}[NS]{neutron star}
\acrodef{GW}[GW]{gravitational wave}
\acrodef{LMXB}[LMXB]{low-mass X-ray binary}
\acrodef{BBH}[BBH]{binary black hole}
\acrodefplural{LMXB}[LMXBs]{low-mass X-ray binaries}
\acrodef{AMXP}[AMXP]{accreting millisecond X-ray pulsar}
\acrodef{EM}[EM]{Electromagnetic}
\acrodef{CW}[CW]{continuous wave}
\acrodef{PSD}[PSD]{power spectral density}
\acrodef{SNR}[S/N]{signal-to-noise ratio}
\acrodef{CPU}[CPU]{central processing unit}
\acrodef{SFT}[SFT]{short Fourier transform}
\acrodef{LLO}[LLO]{LIGO Livingston Observatory}
\acrodef{LHO}[LHO]{LIGO Hanford Observatory}
\acrodef{IFO}[IFO]{interferometer}
\acrodef{GPS}[GPS]{Global Positioning System}
\acrodef{ScoX1}[Sco~X-1]{Scorpius~X-1}
\acrodef{EoS}[EoS]{equation of state}
\acrodef{SSB}[SSB]{Solar system barycenter}
\acrodefplural{EoS}{equations of state}
\acrodef{HMM}[HMM]{hidden Markov model}
\acrodef{LVK}{LIGO-Virgo-KAGRA}
\newcommand{\PEGSi}{PEGS~I}
\newcommand{\PEGSii}{PEGS~II}
\newcommand{\PEGSiii}{PEGS~III}
\newcommand{\PEGSiv}{PEGS~IV}
\preprint{\dcc}

\section{Introduction}
\label{s:intro}

\Acp{NS} in \acp{LMXB} are prime targets for continuous \ac{GW}
searches with the network of current and near-future ground-based
audio band detectors, because of arguments that they will soon be detectable
\citep{Riles2023,Owen2026_Review,Owen2026_Future}.
In these systems, asymmetries in the \ac{NS}'s mass
distribution, sustained by elastic and magnetic stresses \citep{UCB2000,
  Vigelius_2009, Priymak_2011, Singh2020, Osborne2020,
  pagliaro2025}, can emit a persistent, quasimonochromatic \ac{GW}
signal as the \ac{NS} rotates.  Magnetically channeled accretion
provides an additional mechanism to form such a mountain \citep{MP05}.
Accretion
can also spin the \ac{NS} up to a point where gravitational radiation
reaction may drive the growth of \ac{NS} oscillations, such as $r$-modes
\citep{Owen1998, Lindblom1998, Levin_1999, Rezzolla_2000, dong2025},
which emit continuous \ac{GW} signals \citep{Riles2023}.

The \ac{GW} torque may offer the solution to an astrophysical
conundrum. There appears to be a sharp observed cut-off in the spin
frequency ($\nu_s$) distribution of \acp{NS} in \acp{LMXB} at
$\nu_s\approx 750\un{Hz}$ \citep{Chak03, Patruno17}, significantly
below the theoretical breakup frequency \citep{cook1994, Haskell18}.
Although there are uncertainties when modelling the spin-up accretion
torque \citep{Suvorov21, PW21}, which may explain this observation
\citep{Patruno12, Ertan21}, it has also been suggested that the
spin-down \ac{GW} torque leads to an equilibrium at high frequencies,
which naturally explains the observed $\nu_s$ distribution
\citep{Papaloizou1978,Wagoner1984,Bildsten1998}, notably the
clustering for $500\un{Hz}\lesssim \nu_s \lesssim 750\un{Hz}$
\citep{Patruno17, Gittins19}.  In such a scenario, there is a natural
correlation between the observed X-ray flux and \ac{GW} strain, as a
higher accretion rate exerts a stronger spin-up torque which implies a
higher \ac{GW} torque at
equilibrium.

\ac{ScoX1}, the most luminous \ac{LMXB}, which is presumed to
consist of an \ac{NS} of mass $\approx1.4M_{\odot}$ in a binary orbit
with a companion star of mass $\approx0.4M_{\odot}$
\citep{Steeghs2002_ScoX1}, is therefore a promising potential source
of \acp{GW}.  In particular, the torque balance model means that searches
for \acp{GW} from \ac{ScoX1} are now entering the realm where a detectable
signal amplitude is not just possible but predicted.

\begin{deluxetable}{l c}
  \tablewidth{\textwidth}
  \tablecaption{Electromagnetically measured parameters of the \ac{LMXB} \ac{ScoX1}.
    \label{t:ScoX1}}
  \tablehead{
    \colhead{Parameter} & \colhead{Value}
  }
  \startdata
  Right ascension\tablenotemark{a}
  & $\LSCRaHr^{\mathrm{h}}\LSCRaMin^{\mathrm{m}}\LSCRaSec^{\mathrm{s}}$
  \\
  Declination\tablenotemark{a}
  & $-\LSCNegDecDeg^{\circ}\LSCNegDecMin'\LSCNegDecSec''$
 \\
  Distance (kpc) & $2.8\pm0.3$
  \\
  orbital inclination $i$\tablenotemark{b}\qquad\qquad\qquad\qquad\qquad
  & $44^\circ\pm6^\circ$
  \\
  $K_1$ (km/s)\tablenotemark{c} & $[40,90]$
  \\
  $\TascThen$ (GPS s)\tablenotemark{d} & $\KillTascGPS\pm\KilldTascGPS$
  \\
  orbital period $\Porb$ (s)& $\KillPorbSec\pm\KilldPorbSec$
  \\
  \enddata
  \tablerefs{\cite{Bradshaw1999_ScoX1,Fomalont2001,Wang2018_PEGS3,Killestein2023_PEGS4}}
  \tablecomments{ Uncertainties are $1\sigma$ unless otherwise stated.
}
\tablenotetext{a}{The sky position [as quoted in
    \citet{LSC2007_S2ScoX1FStat} derived from \citet{Bradshaw1999_ScoX1}] is
    determined to the microarcsecond.  It is treated as
    being known exactly in this paper.}
\tablenotetext{b}{The inclination $i$ of the orbit to the line of
    sight, from observation of radio jets by \citet{Fomalont2001}, is
    not necessarily the same as the inclination angle $\iota$ of the
    \ac{NS}'s spin axis, which determines the polarization of the
    \ac{GW} in \eref{e:GWamps}, although accretion may drive the two
    angles towards each other.}
\tablenotetext{c}{The projected orbital velocity $K_1$ is estimated by
    Doppler tomography measurements and
    Monte Carlo simulations in \citet{Wang2018_PEGS3}, which show $K_1$ to 
    satisfy the weak constraint
    $40\un{km/s}\lesssim K_1\lesssim 90\un{km/s}$.}
\tablenotetext{d}{ The time of ascension $\TascThen$, at which the
    \ac{NS} crosses the ascending node (moving away from the
    observer), measured in the Solar System barycenter, is given by
    \citet{Killestein2023_PEGS4}.  It corresponds to a
    time of
    {\KillTascDayTime}, and can be propagated to other epochs by
    adding an integer multiple of $\Porb$, which results in increased
    uncertainty in $\Tasc$ and correlations between $\Porb$ and
    $\Tasc$; see \fref{f:TPRegions}.}
\end{deluxetable}

\ac{ScoX1} has been the target of numerous \ac{GW} searches
\citep{LSC2007_S2ScoX1FStat, LSC_S4_radiometer, LVC_S5_radiometer,
  2014CWUnknownBinary, LVC2015_S5ScoX1Sideband,
  Meadors2017_S6ScoX1TwoSpect, LVC2017_O1StochRadiometer,
  LVC2019_O2StochRadiometer, LVK2021_O3StochRadiometer,
  LVK2025_O4aStochRadiometer, Amicucci2025_O3ScoX15Vec} by various
methods \citep{JKS1998_FStat, Ballmer2006_radiometer,
  Messenger2007_Sideband, Goetz2011_TwoSpect, Sammut2014,
  Meadors2015_ScoX1TwoSpect, Leaci2015_ScoX1, Singhal2019_LMXB5Vec,
  Mukherjee2023_BinaryWeave}.
The cross-correlation method
\citep{Dhurandhar2007_CrossCorr,Whelan2015_ScoX1CrossCorr} used in the
present work is an extension of the radiometer search
\citep{Ballmer2006_radiometer} to use the signal model of \acp{GW}
from an \ac{LMXB} such as \ac{ScoX1} to correlate Fourier data at
different times.  It has previously been applied to Advanced LIGO data
from the first, second and third \ac{LVK} observing runs to
set the strongest limits so far on \acp{GW} from \ac{ScoX1}
\citep{LVC2017_O1ScoX1CrossCorr,Zhang2021_O2ScoX1CrossCorr,LVK2022_O3ScoX1CrossCorr,Whelan2023_O3ScoX1NewEphem}.
A complementary method of note is the so-called Viterbi search
\citep{Suvorova2016_Viterbi,Suvorova2017_Viterbi2,Melatos2021_Viterbi3},
which uses a hidden Markov model to track spin wandering and has also
been applied to Advanced LIGO data from the first three \ac{LVK}
observing runs
\citep{LVC2017_O1ScoX1Viterbi,LVC2019_O2ScoX1Viterbi,LVK2022_O3ScoX1Viterbi}.

The fourth and most recent observing run of Advanced LIGO
\citep{LVC2015_aLIGOInst,Soni2025_LIGOO4a,Capote2025_LIGOO4},
Advanced Virgo \citep{Acernese2015_aVirgo,Acernese2023_aVirgo} and KAGRA
\citep{Akutsu2021_KAGRAO3}, O4 for short,
began on {\OivaStartDayTime} (GPS
{\OivaStartGPS}) and ended on {\OivciiEndDayTime} (GPS
{\OivciiEndGPS}).
The portion before the commissioning break at {\OivaEndDayTime} (GPS
{\OivaEndGPS}) is named O4a \citep{LVK2026_O4aData}.  In this paper,
we use O4a data from the two LIGO detectors, as Virgo and KAGRA data
were offline during O4a.  We use the calibrated data
\citep{Sun2020_O3aCal,LVK2025_O4Cal} which are limited to times when a
detector was in scientific observing mode \citep{Soni2025_LIGOO4a}.
Transient instrumental glitches degrade the sensitivity, so we use the
channels \texttt{H1:GDS-CALIB\_STRAIN\_CLEAN\_GATED\_G02} and
\texttt{L1:GDS-CALIB\_STRAIN\_CLEAN\_GATED\_G02} which have had the
``self-gating'' procedure \citep{Davis2025_SelfGating} applied to
remove such glitches.

Some of the parameters inferred from electromagnetic observations of
\ac{ScoX1} are summarized in \tref{t:ScoX1}. The orbital eccentricity
is assumed to be small
\citep{Steeghs2002_ScoX1,Wang2018_PEGS3,Killestein2023_PEGS4} and is
neglected in the present analysis.  A non-circular orbit would add two
search parameters, namely the eccentricity and the argument of
periapse \citep{Messenger2011_Metric,Leaci2015_ScoX1,low2025_herx1},
the effects of which are estimated in \aref{a:eccentricity} and
discussed at the end of \sref{s:CrossCorr}.

The remainder of the paper is laid out as follows. \Sref{s:ScoX1}
describes the modelled \ac{GW} signal from \ac{ScoX1}. \Sref{s:CrossCorr}
defines the resampling cross-correlation search method used in this
paper.  \Sref{s:followup} describes the identification and
followup of potential signals. \Sref{s:upperlimits} sets upper
limits on the \ac{GW} amplitude from \ac{ScoX1}. The conclusions are in \sref{s:conclusions}.

\section{Model of Gravitational Waves from Sco~X-1}

\label{s:ScoX1}

The \ac{GW} signal from a rotating \ac{NS} can be modelled as ``plus''
and ``cross'' polarization components $h_{+}(t)=A_{+}\cos[\Phi(t)]$
and $h_{\times}(t)=A_{\times}\sin[\Phi(t)]$.  The signal received at a
detector is \citep{JKS1998_FStat}
\begin{equation}
  h(t) = F_{+}(t) h_{+}(t) + F_{\times}(t) h_{\times(t)} \ ,
\end{equation}
where $F_{+}$ and $F_{\times}$ are antenna response functions for that
detector in the preferred polarization basis associated with the
source, constructed from amplitude modulation coefficients $a(t)$ and
$b(t)$ (which are detector specific and periodic over one sidereal
day) and an unknown polarization angle $\psi$ which is specific to the
orientation of the \ac{NS} \citep{JKS1998_FStat,Prix2007_MLDC}.

If the \ac{NS} can be modelled as a non-precessing rotating triaxial
ellipsoid, so that the \acp{GW} are generated by the
(non-axisymmetric) equatorial ellipticity of the \ac{NS}, the \ac{GW}
frequency will be twice the spin frequency.  If the \acp{GW} are
generated by $r$-modes, rotating current perturbations in the \ac{NS},
the \ac{GW} frequency is $\sim 1.39-1.64\times$ the spin frequency
\citep{Owen2026_Future}.  This adjusts the approximate multiplicative
factor of $4/3$ from the slow Newtonian approximation
\citep{Papaloizou1978} by including equation of state
\citep{Idrisy2015_rmodes} and relativistic \citep{Ghosh2023_rmodes}
effects.
In
the absence of spin evolution, a signal with frequency $f_0$ in the
reference frame of the source has a phase at the detector of
\begin{equation}
  \label{e:Phidef}
  \Phi(t) = 2\pi f_0 \tbin(t) + \Phi_0 \ ,
\end{equation}
where the relationship between the source frame time $\tbin$ and the
detector time $t$ is determined by the motion of the detector relative
to the \ac{SSB} and the motion of the \ac{NS} in its orbit.  The two
polarization amplitudes are\footnote{This is the form when the
  \acp{GW} are driven by a triaxial deformation of the \ac{NS}.  As
  shown in \citet{Owen2010_rmodes}, for \acp{GW} generated by
  $r$-modes the roles of the plus and cross polarizations are
  interchanged, so quantities like $h_0^{\text{eff}}$ defined in
  \eqref{e:h0eff} remain the same.}
\begin{equation}
  \label{e:GWamps}
  A_{+} = h_0 \frac{1+\cos^2\iota}{2}
  \qquad\hbox{and}\qquad
  A_{\times} = h_0 \cos\iota
  \ ,
\end{equation}
where $h_0$ is an intrinsic amplitude describing the strength of the
signal when it reaches the solar system, and $\iota$ is the
inclination of the \ac{NS}'s spin to the line of sight.  (For an
\ac{NS} in a binary, the spin inclination $\iota$ is not necessarily
equal to the inclination $i$ of the binary orbit; see note $b$ in
\tref{t:ScoX1}.)  If $\iota=0^\circ$ or $180^\circ$, one has
$A_{\times}=\pm A_{+}$, and gravitational radiation is circularly
polarized.  If $\iota=90^\circ$, one has $A_{\times}=0$, and the
radiation is linearly polarized.  The general case, elliptical
polarization, has $0<\abs{A_{\times}}<A_{+}$.  Many search methods are
sensitive to the combination
\begin{equation}
  \label{e:h0eff}
  (h_0^{\text{eff}})^2 = \frac{A_{+}^2+A_{\times}^2}{2}
  = h_0^2\,\frac{[(1+\cos^2\iota)/2]^2 + [\cos\iota]^2}{2}
  \ ,
\end{equation}
which is equal to $h_0^2$ for circular polarization and $h_0^2/8$ for
linear polarization \citep{Messenger2015_MDC1}; this was the
convention used in \citet{LVC2017_O1ScoX1CrossCorr} but differs by a
factor of 2.5 from the definition of $(h_0^{\text{eff}})^2$ in
\citet{Whelan2015_ScoX1CrossCorr}.\footnote{Note that
  $(h_0^{\text{eff}})^2=H_0^2/2$, where $H_0$ is the wave amplitude
  used in the 5-vector literature such as
  \citet{Astone2010_5Vec,Amicucci2025_O3ScoX15Vec}.}

\begin{figure*}[tbp]
  \centering
  \noindent\includegraphics[width=0.33\textwidth]{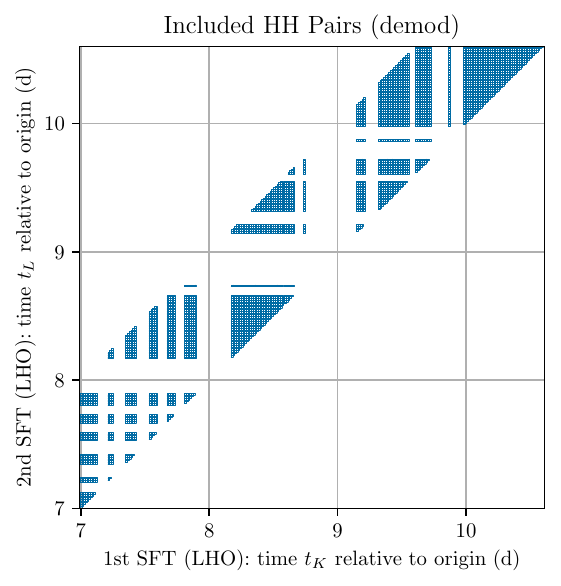}\includegraphics[width=0.33\textwidth]{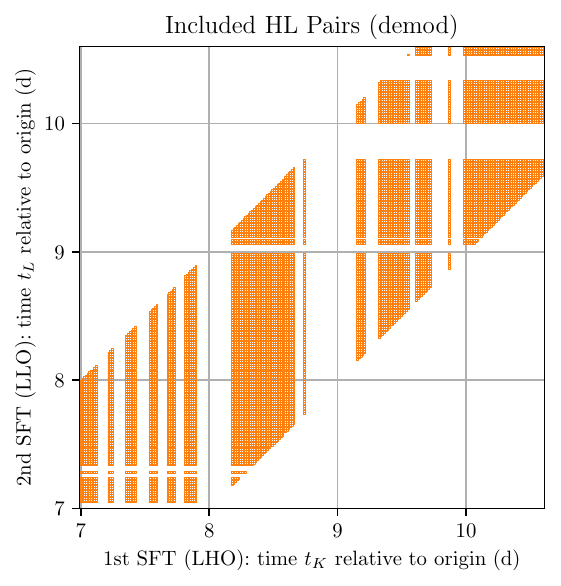}\includegraphics[width=0.33\textwidth]{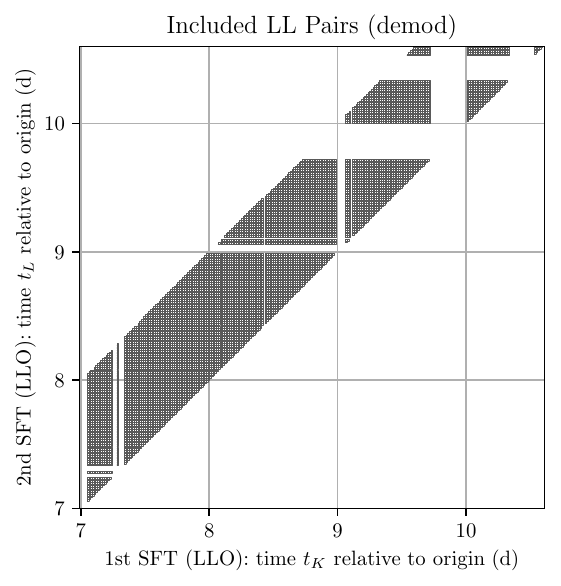}
  \noindent\includegraphics[width=0.33\textwidth]{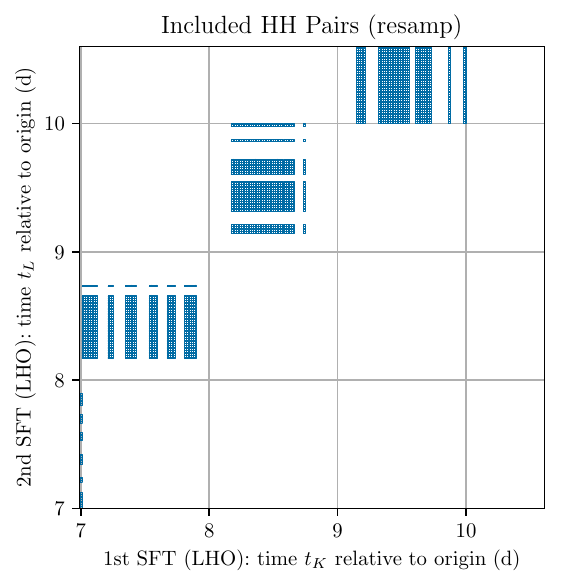}\includegraphics[width=0.33\textwidth]{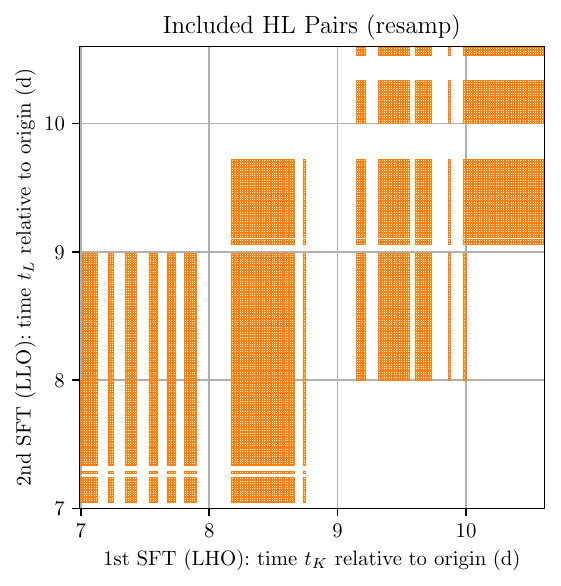}\includegraphics[width=0.33\textwidth]{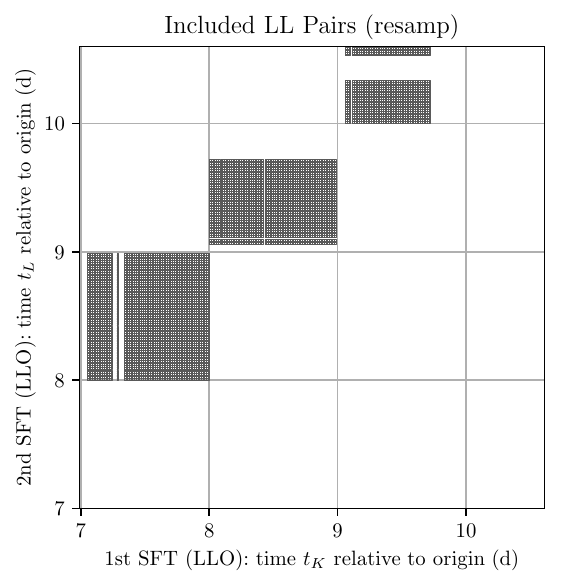}\caption{Representative illustrations of the \acf{SFT} pairs used in
    constructing the cross-correlation statistics.  The left column
    (``HH'') shows the times of pairs where both \acp{SFT} are from
    \acf{LHO}, the middle (``HL'') where the first is from \ac{LHO}
    and the second is from \acf{LLO}, and the right (``LL'') where
    both are from \ac{LLO}.  Note that these plots only show a few
    days of data; the analysis included data from the entire O4a run.
    The horizontal and vertical gaps correspond to times where no
    science-quality data exist for that detector, and so no \acp{SFT}
    were made.  The top row shows the original ``demod'' construction
    used in \eref{e:CCrhodef}: a pair of \acp{SFT} is included in the
    sum if the start times differ by $\Tmax$ or less; in this case
    $\Tmax=24\un{hr}$.  Each pair is only counted once, so for HH and
    LL pairs, we only include pairs with $t_K<t_L$.  This was the
    setup used for the analyses in
    \citet{LVC2017_O1ScoX1CrossCorr,LVK2022_O3ScoX1CrossCorr,Whelan2023_O3ScoX1NewEphem},
    albeit with lower $\Tmax$ values, and is also used for the
    followup described in \sref{s:followup}.  For the resampling
    construction of \citet{Meadors2018_CCResamp}, used in the present
    search, the \ac{SFT} data
    are processed together (with gaps replaced by zeros) into time
    intervals of duration $\Tmax$.  As those are correlated with other
    such intervals that lie within $\Tmax$, the effective set of SFT
    pairs is as shown in the bottom (``resamp'') row.  One effect of
    this is that more HL \ac{SFT} pairs are included in the
    calculation, so that the same nominal $\Tmax$ value corresponds to
    a deeper search.}
  \label{f:pairs}
\end{figure*}

In order to understand the astrophysical relevance of the GW
amplitudes we probe in this paper, a useful benchmark is the
torque-balance level. Suppose that \ac{ScoX1} is in equilibrium, such
that the spin-up torque $N_A$ due to accretion is balanced by a
spin-down \ac{GW} torque
\citep{Papaloizou1978,Wagoner1984,Bildsten1998}.  Let us write
approximately \citep{Pringle72}
\begin{equation}
  N_A=\dot{M}\sqrt{G M\levarm}
  \ ,
\end{equation}
where $\dot{M}$ is the mass accretion rate onto the NS, which we infer
from the X-ray flux $F_X$, $M$ is the mass of the star, $G$ is
Newton's gravitational constant and $\levarm$ is the lever arm,
\textit{i.e.}, the radius at which the accretion torque is applied.
By balancing $N_A$ with the GW torque, we infer the GW amplitude
\citep{Watts2008,Owen2010_rmodes}
\begin{equation}
  \label{e:torquebalgen}
  \begin{split}
  h_0
  \approx
  &\
  5.48\times 10^{-27}
  \left(
    \frac{F_X}{10^{-8}\un{erg}\un{cm}^{-2}\un{s}^{-1}}
  \right)^{1/2}
  \left(
    \frac{f_0}{600\un{Hz}}
  \right)^{-1/2}
  \\
  &\times
  \left(
    \frac{\Rstar}{10\un{km}}
  \right)^{1/2}
  \left(
    \frac{\levarm}{10\un{km}}
  \right)^{1/4}
  \left(
    \frac{M}{1.4M_{\odot}}
  \right)^{1/4}
  \ .
\end{split}
\end{equation}
The usual torque balance benchmark assumes 
$\levarm=\Rstar=10\un{km}$, where $\Rstar$ is the \ac{NS} radius.  If the
magnetic field truncates the accretion disk above
the surface, the lever arm equals the Alfv\'{e}n radius,
$\levarm=r_A$, and $h_0$ implied by
\eref{e:torquebalgen} is larger (e.g.,
\citet{Zhang2021_O2ScoX1CrossCorr,LVK2022_O3ScoX1Viterbi}).
For \ac{ScoX1}, assuming $r=\Rstar$, and using
$F_X=3.9\times10^{-7}\un{erg}\un{cm}^{-2}\un{s}^{-1}$
\citep{Watts2008}, we get
\begin{equation}
  \label{e:torquebal}
  h_0 \approx \htorque
  \left(
    \frac{f_0}{600\un{Hz}}
  \right)^{-1/2}
\end{equation}
for $f_0 = 2\nu_s$.

\Eref{e:torquebal} is an order of magnitude estimate, which we use as
a benchmark to understand if a search probes astrophysically pertinent
portions of the parameter domain.  Much of the physics entering $N_A$
is highly uncertain, and depends on the unknown topology of the
stellar magnetic field, the disk-field coupling, viscous heating in
the disk, efficiency of X-ray emission, or radiation pressure in the
disk \citep{And2005,Patruno12,Haskell2015_mountains,Suvorov21,
  Melatos2023_kalman}.  The amplitude as a function of frequency is
the same whether the \ac{GW} emission mechanism is triaxiality or
$r$-modes \citep{Owen2010_rmodes}.

\section{Resampling Cross-Correlation Search}
\label{s:CrossCorr}

\begin{deluxetable*}{rrrcccccccc}
  \tablewidth{0.9\textwidth}
  \tablecaption{Summary of Numbers of Templates and Candidates.
    \label{t:SearchSummary}}
  \tablehead{
\multicolumn{2}{c}{$f_0$(Hz)} &
&
&
&
&
\multicolumn{5}{c}{candidates at followup level} \\
\colhead{min} &
\colhead{max} &
\colhead{$\Tsft$ (s)} &
\colhead{$\rho$ threshold\tablenotemark{a}} &
\colhead{number of templates} &
\colhead{expected Gauss false alarms\tablenotemark{b}} &
\colhead{cl\tablenotemark{c}} &
\colhead{0\tablenotemark{d}} &
\colhead{1\tablenotemark{e}} &
\colhead{2\tablenotemark{f}} &
\colhead{3\tablenotemark{g}}
}
\startdata
25 &
50 &
1140 &
7.2 &
$4.12\times 10^{11}$ &
0.1 &
68 &
46 &
34 &
18 &
6
\\
50 &
100 &
780 &
7.3 &
$3.24\times 10^{12}$ &
0.5 &
103 &
89 &
84 &
51 &
15
\\
100 &
150 &
660 &
7.5 &
$1.25\times 10^{13}$ &
0.4 &
106 &
92 &
89 &
52 &
14
\\
150 &
200 &
600 &
7.6 &
$3.16\times 10^{13}$ &
0.5 &
134 &
115 &
111 &
70 &
16
\\
\enddata
 \tablecomments{For each range of \ac{GW} frequencies, this table shows the
    \ac{SFT} duration $\Tsft$, the threshold in \ac{SNR} $\rho$ used for
    followup, the total number of templates, and the number of
    candidates at various stages of the process.  (See
    \sref{s:followup} for detailed description of the followup
    procedure.)}
\tablenotetext{a}{This is the threshold for initiating followup.}
\tablenotetext{b}{This is the total number of candidates that would be
    expected in this band due to Gaussian noise, given the number of templates and the
    followup threshold.}
\tablenotetext{c}{This is the actual number of candidates (after
    clustering) which crossed the \ac{SNR} threshold and were followed
    up.}
\tablenotetext{d}{This is the actual number of candidates remaining
    after the ``level 0'' targeted followup.  All of the candidates
    ``missing'' at this stage were vetoed because their \ac{SNR} in
    the targeted search was $<1.645$.}
\tablenotetext{e}{This is the number of candidates remaining after
    refinement.  All of the candidates ``missing'' at this stage have
    been removed by the single-detector veto for unknown lines,
    defined in \sref{s:followup}.}
\tablenotetext{f}{This is the number of candidates remaining after
    each has been followed up with a $\Tmax=4\un{day}$, $4\times$ the
    $\Tmax$ of the level~1 search.  (True signals should approximately
    double their \ac{SNR}; any candidates whose \ac{SNR} goes down
    have been dropped.)  All of the signals present at this stage are
    shown in \fref{f:followupRatio}, which also shows the behavior of
    the search on simulated signals injected in software.}
\tablenotetext{g}{This is the number of candidates remaining after
    $\Tmax$ has been increased to $16\un{day}$.}
\vspace{10pt}
\end{deluxetable*}

\begin{deluxetable}{l c}
  \tablewidth{\textwidth}
  \tablecaption{Parameters used for the Cross-Correlation search.
    \label{t:searchparams}}
  \tablehead{
    \colhead{Parameter} & \colhead{Range}
  }
  \startdata
  $f_0$ (Hz) & $[25,200]$
  \\
  $a\sin i$ (lt-s)\tablenotemark{a} & $[\KillAsiniMin,\KillAsiniMax]$
  \\
  $\TascNow$ (GPS s)\tablenotemark{b}\qquad\qquad\qquad\qquad\qquad
  & $\OivaTascGPS\pm3\times\OivadTascGPS$
  \\
  $\PorbShear$ (s)\tablenotemark{c} & $\KillPorbSec\pm3.3\times\OivadPtildeSec$
  \\
  \enddata
\tablenotetext{a}{The range for the projected
    semimajor axis $a\sin i=K_1\Porb/(2\pi)$ in
    light-seconds was taken from the constraint
    $K_1\in[40,90]\un{km/s}$.
  }
\tablenotetext{b}{This value for the time of ascension $\TascNow$,
    defined in \eref{e:TascNowDef}, has been propagated forward by
    $\OivaNorb$ orbits from the value of $\TascThen$ in
    \tref{t:ScoX1}, and corresponds to a time of {\OivaTascDayTime},
    near the middle of the O4a run.  The increase in uncertainty is due
    to the uncertainty in $\Porb$.  }
\tablenotetext{c}{This is the ``sheared'' period defined in
    \eref{e:Ptildedef}; note that the uncertainty in $\PorbShear$ has
    been reduced compared to the marginal uncertainty in $\Porb$ by
    the same factor by which the uncertainty in $\TascNow$ has been
    increased relative to that for $\TascThen$, as described in
    \citet{Wagner2022_Lattice}.
  }
\end{deluxetable}

\begin{figure}[tbp]
  \centering
  \includegraphics[width=\columnwidth]{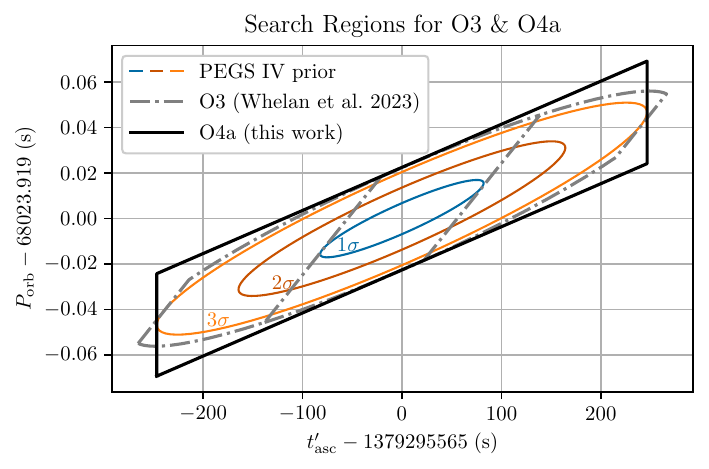}
  \caption{The search region (black parallelogram) for the present
    analysis, defined by $\TascNow$ [\eref{e:TascNowDef}] and
    $\PorbShear$ [\eref{e:Ptildedef}] ranges given in
    \tref{t:searchparams}, expressed in terms of time of ascension
    $\TascNow$ and orbital period $\Porb$.  For comparison, the
    colored ellipses show the (correlated) uncertainties associated
    with the {\PEGSiv} ephemeris \citep{Killestein2023_PEGS4},
    corresponding to $1\sigma$, $2\sigma$, and $3\sigma$ (containing
    $\onesigpct\%$, $\twosigpct\%$, and $\threesigpct\%$ of the prior
    probability, respectively).  The dashed boundaries show the
    truncated elliptical search region of the O3 Cross-Correlation
    analysis in \citet{Whelan2023_O3ScoX1NewEphem}, which also used
    the {\PEGSiv} ephemeris.  Note the dashed diagonal boundaries,
    which are not vertical, because the $\TascNow$ definition used to
    specify them was appropriate for the middle of the O3 run (2019)
    rather than the O4a run ({\OivaTascYear}).  }
  \label{f:TPRegions}
\end{figure}

The cross-correlation (CrossCorr) search method
\citep{Dhurandhar2007_CrossCorr,Whelan2015_ScoX1CrossCorr}, used to
search for \acp{GW} from \ac{ScoX1} in LIGO data from the first three
observing runs of Advanced LIGO, Advanced Virgo, and KAGRA
\citep{LVC2017_O1ScoX1CrossCorr,Zhang2021_O2ScoX1CrossCorr,LVK2022_O3ScoX1CrossCorr,Whelan2023_O3ScoX1NewEphem},
uses the signal model described in \sref{s:ScoX1} to construct an
appropriately weighted statistic $\rho$ including correlations between
data separated by up to a coherence time $\Tmax$.  A reorganization of
the algorithm developed in \citet{Meadors2018_CCResamp} leverages
resampling techniques (converting the data into a time series evenly sampled
in the reference frame of the neutron star)
to speed up the calculation and allow longer
coherence times.  This is primarily effective at lower signal
frequencies, and was used in \citet{Zhang2021_O2ScoX1CrossCorr} to
carry out a search with $\Tmax=19\un{hr}$ for $f_0\in[40,180]\un{Hz}$.

The cross-correlation construction defined in
\citet{Whelan2015_ScoX1CrossCorr}, referred to here as the
``demod'' pipeline, was performed using \acfp{SFT} of length
$\Tsft$.  If the \acp{SFT} are labelled by an index $K$, $L$, etc.,
which encodes the detector and time of the \ac{SFT}, and $z_K$
[defined in eqs.~(2.1), (2.9), (2.19), and~(2.23) of
\citet{Whelan2015_ScoX1CrossCorr}] is an appropriately normalized
combination of the Fourier data at the frequency of interest, the
statistic $\rho$ can be written as
\begin{equation}
  \label{e:CCrhodef}
  \rho = \sum_{KL\in\mc{P}} (W_{KL}z_K^*z_L + W_{KL}^*z_Kz_L^*)
  \ ,
\end{equation}
where $\mc{P}$ is the set of all pairs of \acp{SFT} whose start
times differ by $\Tmax$ or less, as shown in the top row of
\fref{f:pairs}.  The complex weighting factor
\begin{equation}
  W_{KL} \propto \left(
    \widehat{a}_K\widehat{a}_L + \widehat{b}_K\widehat{b}_L
  \right)
  e^{i(\Phi_K-\Phi_L)}
\end{equation}
uses the signal model to determine the expected signal phase
$\Phi_K=\Phi(t_K)$ at the midpoint $t_K$ of \ac{SFT} $K$;
$\widehat{a}_K$ and $\widehat{b}_K$ are noise-weighted antenna
patterns appropriate to the time and detector encoded in $K$.

The resampling technique of \citet{Meadors2018_CCResamp}, which we apply
in this work, reorganizes
the sum in \eref{e:CCrhodef} into a double sum including quantities
like $\widehat{a}_Kz_Ke^{-i\Phi_K}$ and
$\widehat{b}_Kz_Ke^{-i\Phi_K}$.  Since $z_K$ is a combination of
Fourier components, the sum over \acp{SFT} can be written as an
integral over time of $a(t)x(t)e^{-i\Phi(t)}$ or
$b(t)x(t)e^{-i\Phi(t)}$.  Since the signal phase [\eref{e:Phidef}]
$\Phi(t)$ is periodic in the source frame time $\tbin(t)$, resampling
(i.e., reconstructing a time series which is evenly sampled in the
source frame time $\tbin$) turns the time integral within each
\ac{SFT} and the sum over \acp{SFT} into a Fourier transform,
which also replaces the loop over signal frequency $f_0$.  Since the
\ac{SFT} durations are limited by Doppler acceleration, the
calculation is made more efficient by changing the time baseline for
this computation from $\Tsft$ to $\Tmax$ (after resampling and
accounting for the time dependence of the antenna patterns).  This
means that the data from each instrument are divided into intervals of
duration $\Tmax$, and all data lying in intervals separated by $\Tmax$
or less are combined in the calculation of the statistic, as
illustrated in the bottom row of \fref{f:pairs}.

Unlike in the O1 and O3 analyses of
\citet{LVC2017_O1ScoX1CrossCorr,LVK2022_O3ScoX1CrossCorr,Whelan2023_O3ScoX1NewEphem},
which used a range of $\Tmax$ values from $\OiiiTmaxMin\un{s}$ to
$\OiTmaxMax\un{s}$ to tune sensitivity against computing cost, the
current search uses a fixed value of
$\Tmax=24\un{hr}=\OivaTmax\un{s}$, just as the resampling search of
\citet{Zhang2021_O2ScoX1CrossCorr} used a $\Tmax$ of
$19\un{hr}=\OiiTmax\un{s}$.  The \ac{SFT} duration $\Tsft$ is limited
by the Doppler acceleration of the orbit of \ac{ScoX1}, which would
otherwise cause the \ac{GW} signal to depart from the monochromatic
assumption during an \ac{SFT}.  Optimal choices for $\Tsft$ as a
function of signal frequency $f_0$ were determined in
\citet{Whelan2015_ScoX1CrossCorr} and used in
\citet{LVC2017_O1ScoX1CrossCorr,LVK2022_O3ScoX1CrossCorr,Whelan2023_O3ScoX1NewEphem}.
However, in preparation for the present analysis, two errors were
discovered in Figure~4 of \citet{Whelan2015_ScoX1CrossCorr}.  Most
significantly, the assumed projected orbital velocity was taken from
the $a\sin i$ estimates appearing in e.g.,
\citet{LSC2007_S2ScoX1FStat,Watts2008}, which assumed a one-sigma
uncertainty of $a\sin i=1.44\pm0.18\un{lt-s}$ rather than the broader range
of $a\sin i\in[\KillAsiniMin,\KillAsiniMax]\un{lt-s}$
\citep{Wang2018_PEGS3,Killestein2023_PEGS4} shown in \tref{t:ScoX1}.  A
computational error in the opposite direction somewhat mitigated this,
but the end result is that slightly shorter \acp{SFT} are appropriate.
The $\Tsft$ values used in this search are shown in
\tref{t:SearchSummary}.\footnote{We do not believe this introduced
  serious flaws in previous analyses, based on their success in
  recovering injected simulated signals.}

The resampling procedure allows the computation of detection
statistics for multiple frequencies at once, but must be repeated for
different values of the binary orbital parameters which determine the orbital
Doppler shift.  Each combination of orbital parameters and signal
frequency defines a ``search template''.  The range of signal parameters
covered by the search is summarized in
\tref{t:searchparams}.  The projected semimajor axis of the orbit is
assumed to lie in the range $a\sin i\in[\KillAsiniMin,\KillAsiniMax]$
light-seconds, corresponding to a range in projected orbital velocity
of $[40,90]\un{km/s}$.  The time of ascension $\TascThen$ from
\tref{t:ScoX1} is propagated $\OivaNorb$ orbits
(i.e., from {\KillTascYear} to {\OivaTascYear})
to bring it to roughly
the middle of O4a, defining
\begin{equation}
  \label{e:TascNowDef}
  \TascNow = \TascThen + \OivaNorb\times \Porb
  \ .
\end{equation}
The most likely $\TascNow$ is {\OivaTascDayTime} (GPS {\OivaTascGPS}),
and the search covered a range of
$\pm 3\times\OivadTascGPS \un{s}$ centered on that value.  To search
over the range of orbital period $\Porb$ values plausible for each
$\TascNow$, we define the ``sheared'' orbital period
\begin{equation}
  \label{e:Ptildedef}
  \PorbShear = \Porb - \OivadPorbdTascShear (\TascNow-\OivaTascGPS)
\end{equation}
as in \citet{Wagner2022_Lattice}.  The {\coord}s $\TascNow$ and
$\PorbShear$ are approximately uncorrelated both in the parameter
space metric of the search and in the astrophysical prior
distribution.  Note that rather than defining an elliptical region in
$\TascNow$ and $\PorbShear$ as in
\citet{LVK2022_O3ScoX1CrossCorr,Whelan2023_O3ScoX1NewEphem}, we search over a range of $\PorbShear$ values $\pm 3.3\times\OivadPtildeSec \un{s}$
centered on $\KillPorbSec\un{s}$.
The choice of $3$ and $3.3$ times the one-sigma uncertainty in
$\TascNow$ and $\PorbShear$, respectively, means the search region
contains $\tpPct\%$ of the prior probability.
For
reference, in \fref{f:TPRegions} we show this region in the {\coord}s
$(\TascNow,\Porb)$, along with the search regions used in
\citet{Whelan2023_O3ScoX1NewEphem} and the uncertainty ellipses of the
{\PEGSiv} ephemeris \citep{Killestein2023_PEGS4}, propagated forward
in time to the epoch considered in this analysis.

Rather than the $A_n^*$ lattice defined in
\citet{Wagner2022_Lattice,Wette2014_Lattice} and used in
\citet{LVK2022_O3ScoX1CrossCorr,Whelan2023_O3ScoX1NewEphem}, the
present analysis uses a rectangular grid as in
\citet{LVC2017_O1ScoX1CrossCorr,Zhang2021_O2ScoX1CrossCorr}.  The
metric mismatch \citep{Whelan2015_ScoX1CrossCorr} \emph{between
  templates} was taken to be $\mismatchMax$, which corresponds, for a
four-dimensional grid, to a maximum mismatch between any point in the
parameter space and the nearest template which is also no more than
$\mismatchMax$.  Future development is planned to implement the more
efficient $A_n^*$ lattice within the resampling CrossCorr pipeline.

\ac{ScoX1} may undergo stochastic fluctuations in $f_0 $,
known as spin wandering
\citep{Bildsten1997_Spinwander,Mukherjee2018_Spinwander}.  One can set
bounds on the loss of \ac{SNR} within the framework of an idealized
model, in which $f_0$ undergoes a net spin-up or spin-down of
magnitude $\fdotdrift$, fluctuating on a time scale $\Tdrift$
\citep{Sammut2014,Messenger2015_MDC1,Whelan2015_ScoX1CrossCorr,LVC2017_O1ScoX1CrossCorr,Carlin2025_Spinwander}.
For O4a, whose duration from start to end is $\Trun=\OivaTrun \un{s}$,
and coherence time $\Tmax= \OivaTmax\un{s}$, the expected loss of
\ac{SNR} is, applying eq.~(5.9) of \citet{Whelan2015_ScoX1CrossCorr},
\begin{multline}
  \label{e:SWloss}
  \frac{\ev{\rho}^{\text{ideal}}-\ev{\rho}}{\ev{\rho}^{\text{ideal}}}
  \\
  \approx
  \OivaSWLoss
  \left(\frac{\Tdrift}{10^6\un{s}}\right)
  \left(\frac{\fdotdrift}{10^{-12}\un{Hz/s}}\right)^2
  \left(\frac{\Tmax}{\OivaTmax\un{s}}\right)^2
  \ .
\end{multline}
This gives some indication that spin wandering could be an issue for
this longer coherence time.
\citet{Carlin2025_Spinwander}, which assumes a somewhat higher level of
spin wandering, predicts an effect consistent with the scaling in
\eref{e:SWloss}.
However, as noted in \citet{Zhang2021_O2ScoX1CrossCorr} and
\citet{LVK2022_O3ScoX1CrossCorr}, the predictions
of \citet{Mukherjee2018_Spinwander} based on the variability of $F_X$
imply considerably less spin wandering than the {\naive}
$\fdotdrift$--$\Tdrift$ model, whose fiducial parameters are taken from
\citet{Messenger2015_MDC1}, so this search is still likely to be sensitive to a realistic signal.

Finally, the present search assumes a circular orbit for \ac{ScoX1},
in keeping with the ephemeris of \citet{Killestein2023_PEGS4}.  As
computed in \eref{e:eccmismatch} of \aref{a:eccentricity}, the
fractional loss of \ac{SNR} due to a small orbital eccentricity $e$
can be estimated as
\begin{multline}
  \frac{\ev{\rho}^{\text{ideal}}-\ev{\rho}}{\ev{\rho}^{\text{ideal}}}
  \\
  \approx
  \EccLoss
  \left(\frac{f_0}{200\un{Hz}}\right)^2
  \left(\frac{a\sin i}{\KillAsiniMax\un{lt-s}}\right)^2
  \left(\frac{e}{10^{-4}}\right)^2
  \ ,
\end{multline}
from which we conclude it is not an important effect for this search.

\begin{figure}[tbp]
  \centering
  \includegraphics[width=\columnwidth]{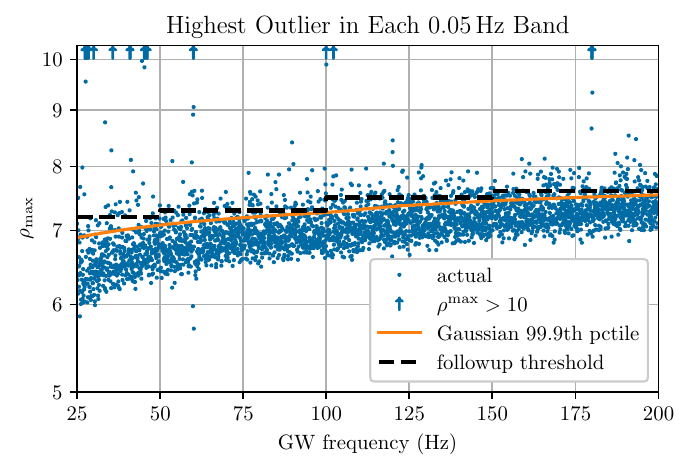}
  \caption{Selection of followup threshold as a function of \ac{GW}
    frequency.  The blue dots represent the highest \ac{SNR} $\rhomax$
    in each $0.05\un{Hz}$ band.  If the data contained no signal and
    only Gaussian noise, each template in the parameter space would
    have some chance of producing a statistic value exceeding a given
    threshold.
Using the number of templates in each $0.05\un{Hz}$ band, we can
    estimate the expected distribution of the highest \ac{SNR} in the
    band.  The solid orange curve shows the 99.9th percentile of this
    distribution, i.e., the level at which we would expect $10^{-3}$ false
    alarms per $0.05\un{Hz}$, or one per $50\un{Hz}$.  Due to the
    non-Gaussian nature of the tail of the distribution of $\rho$
, the actual number
    of outliers is considerably higher.  To keep the followup volume
    manageable, we set the thresholds indicated by the black dashed
    line and shown in \tref{t:SearchSummary}.}
\label{f:threshold}
\end{figure}

\section{Candidates and Followup}
\label{s:followup}

The detection statistic $\rho$ is normalized to have zero mean and
(for low signal amplitudes) unit variance in Gaussian noise.  A
{\naive} estimate of the expected background would assume each search
template to represent an independent Gaussian random number, and set a
threshold based on a desired number of outliers to follow up, as was
done in
\citet{LVC2017_O1ScoX1CrossCorr,LVK2022_O3ScoX1CrossCorr,Whelan2023_O3ScoX1NewEphem}.\footnote{A more sophisticated way of estimating this distribution would be the method of \citet{Tenorio2022_distromax}.}
The present search, due to the longer coherence time, uses many more
templates for a given frequency range (see \tref{t:SearchSummary}),
and therefore requires a higher threshold to maintain a desirable
outlier rate.  This is compounded by the non-Gaussianity of the tail
of the distribution of $\rho$; as discussed in Appendix~B of
\citet{Whelan2015_ScoX1CrossCorr} this can occur even if the instrumental
noise is Gaussian.  As shown in \fref{f:threshold}, we
set our followup threshold at $\rhothresha$ for
$\fmina\un{Hz}<f_0<\fmaxa\un{Hz}$, $\rhothreshb$ for
$\fminb\un{Hz}<f_0<\fmaxb\un{Hz}$, $\rhothreshc$ for
$\fminc\un{Hz}<f_0<\fmaxc\un{Hz}$, and $\rhothreshd$ for
$\fmind\un{Hz}<f_0<\fmaxd\un{Hz}$.  \Tref{t:SearchSummary} shows the
resulting expected numbers of false alarms in each \ac{GW} frequency
range.

\begin{figure}[tbp]
  \centering
  \includegraphics[width=\columnwidth]{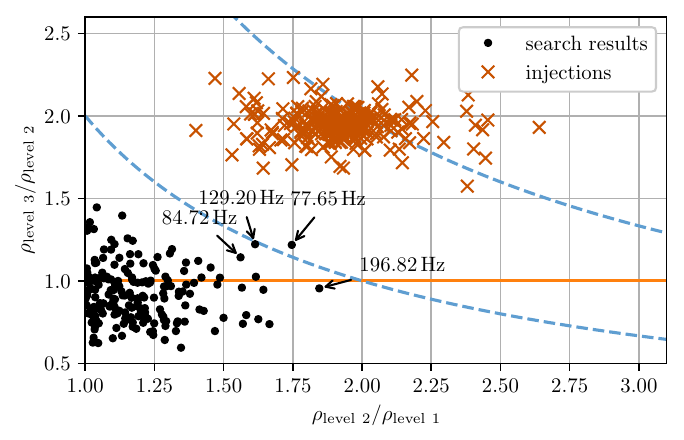}
  \caption{Ratios of followup statistics for candidates from the search,
    and from
    simulated signals.  This plot shows all of the candidates that
    survived level~2 of followup (see \sref{s:followup} and
    \tref{t:SearchSummary}), both from the main search and from the
    analysis of the simulated signal injections described in
    \sref{s:upperlimits}.  It shows the ratios of the \ac{SNR} $\rho$
    after followup level~1 (a refined demod CrossCorr search with
    $\Tmax=24\un{hr}$), level~2 ($\Tmax=4\un{day}$), and level~3
    ($\Tmax=16\un{day}$).  The blue dashed lines are at constant
    values of $\rho_{\text{level 3}}/\rho_{\text{level 1}}$ equal to
    $2$ and $4$, the latter being the theoretically expected
    value.  There are no points with
    $\rho_{\text{level 2}}/\rho_{\text{level 1}}<1$, because those
    candidates do not survive level~2 followup and are therefore not
    subjected to level~3 followup.  From the construction of the
    statistic in \citet{Whelan2015_ScoX1CrossCorr}, the {\naive}
    expectation is that the \ac{SNR} will roughly double each time
    $\Tmax$ is quadrupled.  We see that the followups of injections
    (red crosses) are clustered around this relationship, while the
    followups of search candidates (black dots) are not.  The
    four candidates which increase their \ac{SNR} by the largest factor
    from level 1 to level 3 are indicated,
    labelled by their frequencies, and considered in
    \tref{t:Outliers}.}
  \label{f:followupRatio}
\end{figure}

\begin{deluxetable*}{rcccccccc}
  \tablewidth{0.9\textwidth}
  \tablecaption{Summary of Loudest Outliers.
    \label{t:Outliers}}
  \tablehead{
\colhead{$f_0$(Hz)} &
\colhead{$\rho_{\text{resamp}}$\tablenotemark{a}} &
\colhead{$\rho_{\text{lvl\,1}}$\tablenotemark{b}} &
\colhead{$\rho_{\text{lvl\,2}}$\tablenotemark{c}} &
\colhead{$\rho_{\text{lvl\,3}}$\tablenotemark{c}} &
\colhead{$h_{0,\text{resamp}}^{\text{eff}}$\tablenotemark{d}} &
\colhead{$h_{0,\text{lvl\,1}}^{\text{eff}}$\tablenotemark{e}} &
\colhead{$h_{0,\text{lvl\,2}}^{\text{eff}}$\tablenotemark{e}} &
\colhead{$h_{0,\text{lvl\,3}}^{\text{eff}}$\tablenotemark{e}}
}
\startdata
77.65 &
7.39 &
3.81 &
6.66 &
8.11 &
$2.31\times 10^{-26}$ &
$1.80\times 10^{-26}$ &
$1.71\times 10^{-26}$ &
$1.34\times 10^{-26}$
\\
84.72 &
7.74 &
4.76 &
7.43 &
8.48 &
$2.27\times 10^{-26}$ &
$2.09\times 10^{-26}$ &
$1.87\times 10^{-26}$ &
$1.43\times 10^{-26}$
\\
129.20 &
7.84 &
5.93 &
9.58 &
11.70 &
$1.89\times 10^{-26}$ &
$1.80\times 10^{-26}$ &
$1.64\times 10^{-26}$ &
$1.29\times 10^{-26}$
\\
196.82 &
7.65 &
3.91 &
7.22 &
6.89 &
$1.63\times 10^{-26}$ &
$1.28\times 10^{-26}$ &
$1.24\times 10^{-26}$ &
$8.65\times 10^{-27}$
\\
\enddata
 \tablecomments{These are the outliers in \fref{f:followupRatio} for
    which the \ac{SNR} increased by a factor of more than
    {\outlierratio} when the coherence time was increased by a factor
    of 16 from level 1 to level 3.  In all but one of the cases, the
    \ac{SNR} dropped more than expected in the switch from the
    original resamp search to the demod followup, as reflected in the
    reduction of the estimated $h_0^{\text{eff}}$ associated with the
    outlier.  In those cases, the increase in \ac{SNR} from level 1 to
    level 3 still resulted in $\rho_{\text{lvl\,3}}<10$.  The one
    exception was the signal at $\outlierf\un{Hz}$, which, while
    seeing a reduction in estimated $h_0^{\text{eff}}$ at each stage,
    still ended up with an \ac{SNR} of
    $\rho_{\text{lvl\,3}}=\outlierrhoiii$.  For comparison, the lowest
    $\rho_{\text{lvl\,3}}$ for injections which
    survived the veto procedure was $\injminrhoiii$.}
\tablenotetext{a}{The \ac{SNR} of the candidate in the initial
    resampling search with $\Tmax=24\un{hr}$.}
\tablenotetext{b}{The \ac{SNR} of the candidate in the ``level 1''
    refinement followup.  Note that while this also has
    $\Tmax=24\un{hr}$, it is less sensitive than the initial search
    because the implications of a $\Tmax$ value are different in the
    resampling and demod searches, as shown in \fref{f:pairs}.}
\tablenotetext{c}{The \ac{SNR} of the candidate in the ``level 2''
    and ``level 3'' followup searches with $\Tmax=4\un{day}$ and
    $16\un{day}$, respectively}
\tablenotetext{d}{The estimated $h_0^{\text{eff}}$ \eqref{e:h0eff}
    of the candidate in the original resampling search, estimated from
    the \ac{SNR} and the standard sensitivity $\curlyrho$ defined in
    e.g., eq.~(7) of \citet{LVC2017_O1ScoX1CrossCorr}, recalling that a
    true signal has expected \ac{SNR}
    $\ev{\rho}=(h_0^{\text{eff}})^2\curlyrho$.}
\tablenotetext{e}{The estimated $h_0^{\text{eff}}$ of the candidate
    in the corresponding level of followup.}
\vspace{10pt}
\end{deluxetable*}

For candidates exceeding the followup threshold, we applied a modified
version of the hierarchical strategy detailed in
\citet{LVC2017_O1ScoX1CrossCorr,LVK2022_O3ScoX1CrossCorr}.  In
particular, we used the non-resampling ``demod'' version of the
CrossCorr pipeline in the followup stages.  This version of the
pipeline automatically zeros out frequencies associated with known
instrumental disturbances (``lines'') \citep{Goetz2025_Lines},
reducing the number of outliers reaching the later stages of followup.
The steps were as follows:
\begin{itemize}
\item Candidates were iteratively ``clustered'' together in \ac{GW}
  frequency, with all templates within $0.01\un{Hz}$ of the highest
  \ac{SNR} being represented by the parameters of this peak, and the
  process repeated on the remaining templates until no more remained
  above the threshold.  Note that choosing too low a followup
  threshold would result in effectively every $0.02\un{Hz}$ band
  containing a candidate.
\item A targeted search using the demod pipeline with
  $\Tmax=24\un{hr}$ was performed on every clustered candidate.  The
  primary purpose of this step is to determine the parameter space
  metric for later stages of followup, but it is also used to
  eliminate candidates arising from known instrumental lines by
  vetoing any candidate with $\rho<1.645$.  Note that the \ac{SNR} of a
  candidate is expected to go down slightly at this stage because
  $\Tmax=24\un{hr}$ corresponds to less correlated data in the demod
  search than in the original resampling search, as shown in
  \fref{f:pairs}.  Candidates surviving this veto are known as the
  ``level~0'' results.
\item A ``refinement'' search was performed on each level~0 candidate,
  with $\Tmax=24\un{hr}$ and a resolution $\sim 3\times$ as fine as
  the $0.25$ mismatch grid implied by the level~0 metric.  This and
  later stages of followup were run on a rectangular grid in the
  ``sheared'' parameters $(f_0,a\sin i,\TascNow,\PorbShear)$.  For the
  refinement stage, the grid contained $13$ points in each of the
  $(f_0,a\sin i,\TascNow)$ directions, and between $5$ and $13$ points
  as necessary to cover the prior range of $\PorbShear$.  To deal with
  the effects of unknown narrow-band features (``lines'') present in a
  single detector, we computed a detection statistic using only data
  from the \ac{LLO} detector and another using only \ac{LHO} data.  If
  either of these exceeded the detection statistic constructed from
  all the data, we vetoed the candidate as a likely instrumental
  artifact.  Candidates from the refinement search that survive this
  veto are known as the ``level~1'' results.
\item Two successive rounds of followup were performed, starting with
  the level~1 candidates.  At each stage, the coherence time $\Tmax$
  was quadrupled from the previous stage, and the density of templates
  in the $f_0$ direction was increased by a factor of 3.  The grid
  used was $13\times5\times5\times5$ in
  $(f_0,a\sin i,\TascNow,\PorbShear)$, centered on the peak of the
  previous level's results.  No further refinement was
  necessary in the orbital parameter directions because the orbital
  metric saturates for $\Tmax>\Porb\approx{\KillPorbHr\un{hr}}$.
  Candidates that increased their \ac{SNR} from the previous level
  were known as the ``level~2'' ($\Tmax=4\un{day}$) and ``level~3''
  ($\Tmax=16\un{day}$) results.
\end{itemize}
A total of {\numUnvetoed} candidates survive level~3 of followup.  To
determine whether any of them represent convincing detection
candidates, we examine the factor by which each increases its \ac{SNR}
when the coherence time is quadrupled from level~1 ($\Tmax=24\un{hr}$)
to level~2 ($\Tmax=4\un{day}$) and again from level~2 to level~3
($\Tmax=16\un{day}$).  We plot this in \fref{f:followupRatio} for all
of the candidates surviving level~2 (the level~3 followup is not run
for candidates which fail level~2), and also for the simulated signal
injections described in \sref{s:upperlimits}.  We would {\naive}ly
expect a real signal to double its \ac{SNR} between level~1 and
level~2, and again between level~2 and level~3.  All of the candidates
arising from the search fall
well short of this.  For comparison, we also applied the followup
procedure to the {\numInjFollowup} injected signals (out of
{\numInjTotal}) that produced $\rho$ values above their respective
thresholds.  Of these, {\numInjVetoedi} were vetoed at level~1,
{\numInjVetoedii} at level~2 and {\numInjVetoediii} at level~3, all
because of the single-detector unknown-line veto.  The remaining
{\numInjUnvetoed} survived all the levels of followup, and
approximately doubled their \ac{SNR} at each stage, as shown in
\fref{f:followupRatio}

We give details of some of the most interesting
outliers in \tref{t:Outliers}.
The most intriguing outlier from the search is at a frequency of
$\outlierf\un{Hz}$.  As shown in \tref{t:Outliers}, it has an initial
$\rho$ of $\outlierrho$ in the original resampling search.  The
targeted level~0 and refined level~1 followups (both with
$\Tmax=24\un{hr}$) produced $\rho_{\text{level 0}}=\outlierrhoo$ and
$\rho_{\text{level 1}}=\outlierrhoi$. This reduction in
\ac{SNR} is expected because of the smaller number of \ac{SFT} pairs
in the demod search at a given $\Tmax$, and indeed the estimated
$h_0^{\text{eff}}$ decreases negligibly between the original search
and the level~1 followup.  At level~2 ($\Tmax=4\un{day}$) the signal
increased its \ac{SNR} by a factor of $\outlierratiotwoone$ to
$\outlierrhoii$ and at level~3 ($\Tmax=16\un{day}$) it increased by a
further factor of $\outlierratiotretwo$ to $\outlierrhoiii$, the
highest followup \ac{SNR} of any of our candidates.  This is of
particular interest because a signal with spin wandering might be
expected to have a lower \ac{SNR} than predicted in searches with
longer coherence times of $\Tmax\gtrsim 10\un{day}$.

To check whether the $\outlierf\un{Hz}$ could be due to a signal whose
frequency was stochastically ``wandering'', an analysis was performed
with the Viterbi/HMM pipeline
\citep{Suvorova2016_Viterbi,Suvorova2017_Viterbi2} over a narrow range
of frequency and orbital parameters centered on those of the outlier.
This search also found a feature at the parameters of this
candidate, which survived its known lines \citep{Covas2018,
  Goetz2026_Lines}, single interferometer
\citep{LVK2022_O3ScoX1Viterbi} and Doppler modulation-off
\citep{Jones2022_PSF} vetoes.  However, the most likely path for
the wandering signal frequency oscillated between two frequencies
separated by less than {$2\times 10^{-7}\un{Hz}$}.  At the
resolution of the final CrossCorr followup step, a frequency offset of
{$\pm 10^{-7}\un{Hz}$} would result in a loss of \ac{SNR} of
less than {13\%}, so the difference in behavior between the
candidate and simulated signals does not seem to be explained by spin
wandering.

As an independent check of the $\outlierf\un{Hz}$ candidate,
we applied the resampling CrossCorr search
with $\Tmax=24\un{hr}$ to the corresponding region of parameter space
in data from the second part of the fourth observing run (O4b), which
began on {\OivbStartDayTime} (GPS
{\OivbStartGPS}) and ended on {\OivbEndDayTime} (GPS
{\OivbEndGPS}),
but no
signal was seen with {$\rho>6.1$}.  A true persistent signal at
these parameters should have been seen with {$\rho\gtrsim 8.2$},
given the greater duration and sensitivity of the O4b dataset compared
to O4a.  We conclude that this feature in the O4a data, while
suggestive, does not appear to be associated with a gravitational wave
signal.

\begin{figure*}[tbp]
  \centering
  \includegraphics[width=0.45\textwidth]{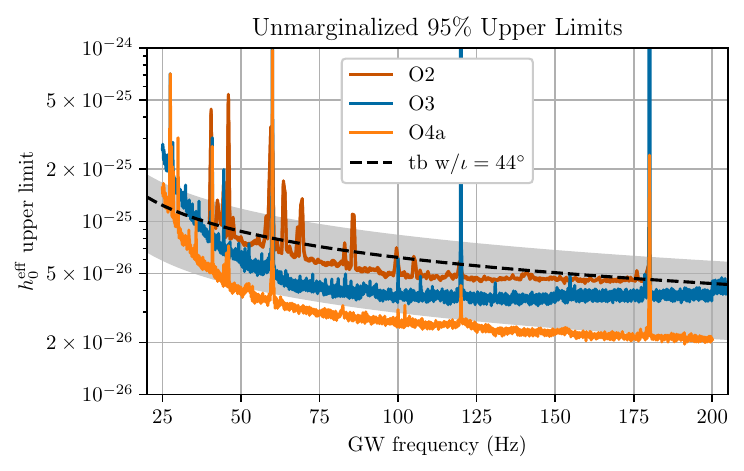}
  \includegraphics[width=0.45\textwidth]{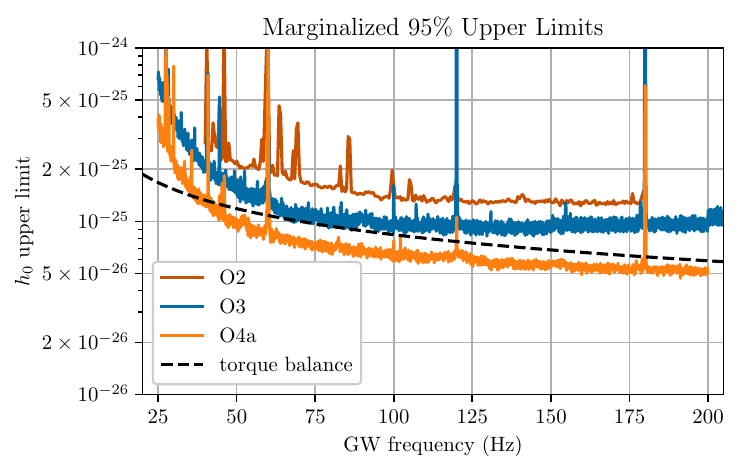}
  \caption{Upper limits from cross-correlation searches for \ac{ScoX1}
    in Advanced LIGO
    data.
Left: The upper limit shown is on $h_0^{\text{eff}}$, defined in
    \eref{e:h0eff}.  This is equivalent to the upper limit on $h_0$
    assuming circular polarization. The ``O4a'' curve shows the results of the present resampling
    CrossCorr search.  ``O3'' is the CrossCorr search
    \citep{LVK2022_O3ScoX1CrossCorr}, and ``O2'' is the resampling
    CrossCorr search of \citet{Zhang2021_O2ScoX1CrossCorr}.  For
    comparisons among O3 searches with different methods, see
    \citet{LVK2022_O3ScoX1CrossCorr}.
The grey shaded region corresponds to the range of amplitudes corresponding to the nominal expected level assuming torque balance
    [\eref{e:torquebal}], which depends on the unknown inclination angle.
    The dashed line (labelled as ``tb w/$\iota=44^{\circ}$'')
    corresponds to the assumption that the \ac{NS} spin is
    aligned to the most likely orbital angular momentum, and
    $\iota\approx i\approx 44^\circ$.  (See \tref{t:ScoX1}.)
Right: marginalized upper limits on $h_0$ assuming an isotropic prior on the neutron star spin inclination $\iota$, with the dashed line indicating the level predicted by torque balance.
(Note that this marginalized upper
    limit is dominated by linear polarization, and
    so is a factor of almost $\sqrt{8}$ higher than the unmarginalized upper limit.)
The present search has, for the first time, set an upper limit below the torque balance prediction at \emph{any} inclination angle for $50\un{Hz}<f_0<200\un{Hz}$.
}
  \label{f:UL}
\end{figure*}

\begin{figure}[tbp]
  \centering
  \includegraphics[width=\columnwidth]{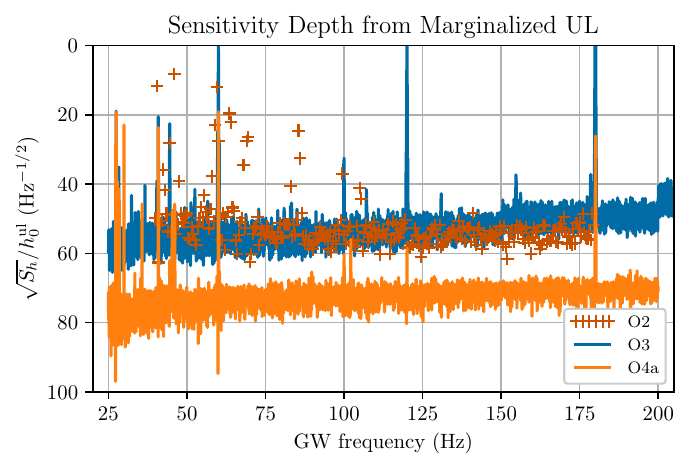}
  \caption{The marginalized upper limits from \fref{f:UL}, expressed
    in terms of the quantity known as sensitivity depth
    \citep{Behnke2015_SensDepth,Dreissigacker2018_SensDepth}
    $\mc{D}=\sqrt{S(f)}/h_0^{\text{ul}}$, which scales out the
    effect of detector sensitivity by using the averaged noise power
    spectral density $S(f)$ for the observing run in question.  Note
    that the (resampling) O2 search of
    \citet{Zhang2021_O2ScoX1CrossCorr} and the (demod) O3 search of
    \citet{LVK2022_O3ScoX1CrossCorr} have similar sensitivity depths
    of around {$50-60\un{Hz}^{-1/2}$}, with the longer
    coherence time of the resampling analysis being offset by the
    longer duration of O3.  The variable choice of $\Tmax$ in the O3
    analysis, designed to trade off computing cost vs scientific
    return, results in a deeper search at low frequencies and
    shallower at high frequencies.  The present O4a resampling
    analysis achieves a sensitivity depth of around
    {$70-75\un{Hz}^{-1/2}$}.}
  \label{f:sensDepth}
\end{figure}

\begin{figure}[tbp]
  \centering
  \includegraphics[width=\columnwidth]{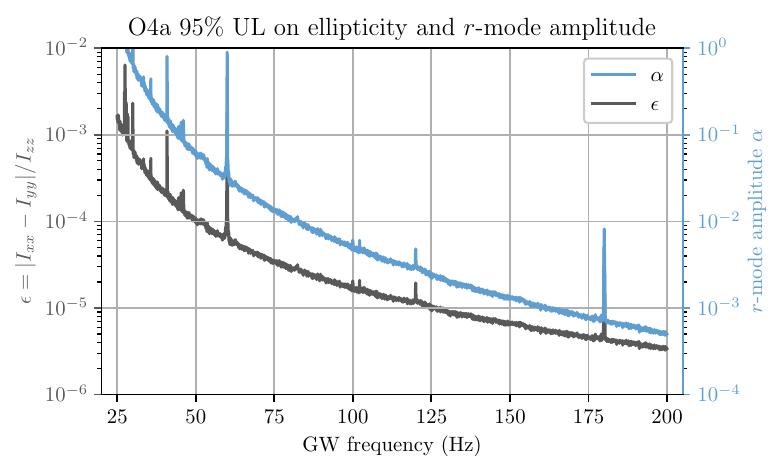}
  \caption{The marginalized upper limits from \fref{f:UL}, converted
    to limits on the ellipticity
    $\epsilon=\frac{\abs{I_{xx}-I_{yy}}}{I_{zz}}$
    \citep{Riles2023,Wette2023_Review} of the \ac{NS} in \ac{ScoX1}
    (left axis scale) or alternatively the $r$-mode amplitude $\alpha$
    (right axis scale).  The ellipticity assumes a moment of inertia of
    $10^{38}\un{kg}\un{m}^2$ and a distance of
    {$2.8\un{kpc}$}.  For \ac{GW} frequencies between
    {$75\un{Hz}$} and {$200\un{Hz}$}, the constraints
    are between {$3\times 10^{-5}$} and
    {$3\times 10^{-6}$}.  At the highest frequencies, these
    are entering the range of possible ellipticities for ordinary
    \ac{NS} matter.  The $r$-mode amplitudes use equation~(24) of
    \citet{Owen2010_rmodes}; the best limits of
    {$\sim 5\times 10^{-4}$} are probably still slightly above
    what one might expect in \ac{ScoX1} (\citet{Owen2026_Future}
    and references therein).}
  \label{f:ellipandrmode}
\end{figure}

\section{Upper Limits and Implications}
\label{s:upperlimits}

In the absence of convincing detection candidates, we set upper limits
on the strength of \acp{GW} from \ac{ScoX1} as a function of \ac{GW}
frequency, using the method described in detail in
\citet{LVC2017_O1ScoX1CrossCorr}.  The starting point is a {\naive}
Bayesian upper limit within each $0.05\un{Hz}$ band, set using the
95th percentile of the posterior on $h_0^2$ or $(h_0^{\text{eff}})^2$
deduced from the highest \ac{SNR} seen in each band.  These were
adjusted by a factor determined through Bayesian logistic regression
on injected simulated signals.

The upper limits are plotted in \fref{f:UL}, and for the first time
set constraints below the standard torque balance limit for all values
of the neutron star spin inclination angle, across a range of
frequencies from $50$ to $200\un{Hz}$.  The most sensitive upper
limits are at the upper end of the frequency band, and approach a
\ac{GW} amplitude of {$5\times 10^{-26}$} marginalized over
inclination angle and {$2\times 10^{-26}$} assuming the most
favorable inclination.  In \fref{f:sensDepth}, we show the
marginalized upper limits in terms of the sensitivity depth
$\mc{D}=\sqrt{S(f)}/h_0^{\text{ul}}$ defined in
\citet{Behnke2015_SensDepth,Dreissigacker2018_SensDepth}, which
separates the inherent sensitivity of the search from the effects
of the reduction of noise level between observing runs.  The present
search achieves a sensitivity depth of
$\mc{D}\sim 70-75\un{Hz}^{-1/2}$, improving upon previous
searches, which had a depth of around $50-60\un{Hz}^{-1/2}$.

In \fref{f:ellipandrmode} we convert the upper limits into limits on
the physical process associated with the \ac{GW} generation.  In the
case of triaxiality, we compute an upper limit on ellipticity
$\epsilon=\frac{\abs{I_{xx}-I_{yy}}}{I_{zz}}$ assuming a gravitational
wave amplitude of \citep{Riles2023,Wette2023_Review}
\begin{equation}
  h_0 = \frac{4\pi^2 G}{c^4 d} (I_{zz})\epsilon
  f_0^2 \ ,
\end{equation}
where we assume the most likely distance $d=2.8\un{kpc}$ from
\tref{t:ScoX1} and a moment of inertia of
$I_{zz}=10^{38}\un{kg}\un{m}^2$.
For $f_0$ between $75$ and $200\un{Hz}$, the upper limits
(marginalized isotropically over inclination angle) are between
$3\times 10^{-5}$ and $3\times 10^{-6}$, probing ellipticities which
are potentially supportable by ordinary matter
\citep{Johnson-McDaniel2013_ellipticity,Gittins2024_Mountains}.
For the case of \acp{GW} driven by $r$-modes, we use equation~(24) of
\citet{Owen2010_rmodes}:
\begin{equation}
  \label{e:rmodeamp}
  \alpha = 0.028 \left(\frac{h_0}{10^{-24}}\right)
  \left(\frac{d}{1\un{kpc}}\right)\left(\frac{100\un{Hz}}{f_0}\right)
  \ .
\end{equation}
Our strongest constraints are at the upper end of the frequency range,
$f_0\approx 200\un{Hz}$ [corresponding to a spin frequency of
$120-150\un{Hz}$ \citep{Owen2026_Future}], and amount to (subject to
all the assumptions that go into \eqref{e:rmodeamp})
$\alpha\lesssim 5\times 10^{-4}$.  This is just within the plausible
range of saturation amplitudes for a newborn \ac{NS}
\citep{Bondarescu2009_rmodes}, but still slightly above what is
expected for an accreting \ac{NS} such as \ac{ScoX1}
\citep{Bondarescu2007_rmodes}.

\label{s:implications}

From an astrophysical point of view, the upper limits from the
  current search allow us to essentially exclude the \ac{GW}-dominated
  torque balance
  scenario in the range between roughly $50$ and $200\un{Hz}$. On the
  one side the torque balance amplitude in \eref{e:torquebal}
  already considers the most stringent upper limit by taking the
  torque arm to correspond to the stellar surface, and on the other it
assumes a very compact neutron star, with mass $M=1.4 M_{\odot}$ and
radius $R=10\un{km}$. Microphysically motivated \acp{EoS}
generally predict less compact stars, and even for softer \ac{EoS}, such as
for example the GR15 \ac{EoS} \citep{GR15} considered by
\citet{LVK2022_O3ScoX1CrossCorr}, one cannot construct a model that
would allow for torque balance. Although it is important to keep in mind that significant
uncertainties are present in torque modelling, and that GW torques may
be active only part of the time \citep{Haskell2015_mountains}, our
results essentially exclude the possibility that \ac{ScoX1} includes an
\ac{NS} in equilibrium between accretion torques and \ac{GW} torques
with a \ac{GW} frequency between approximately $50$ and $200\un{Hz}$.
(This corresponds to a spin frequency $\nu_s$ between $25$ and
$100\un{Hz}$ if triaxiality is the dominant \ac{GW} emission
mechanism, or $\sim 15-20$ and $\sim 120-150\un{Hz}$ if the \ac{GW}s
are driven by $r$-modes.)
The torque balance
scenario remains hypothetically possible for a very compact quark
star, an object which has been conjectured to exist in \acp{LMXB}
\citep{quarkLMXB_2003}. In this case, if for example we consider a
quark star described by the MIT bag model, with a bag constant $B=60\un{MeV/fm}^3$, then stars with masses $M\lesssim 0.5\, M_{\odot}$ (and
corresponding radii $R\lesssim 8\un{km}$) would still be consistent with
the torque balance scenario and the reported upper limits.

\section{Conclusions and Outlook}
\label{s:conclusions}

We have presented the results of the most sensitive search to date for
\acp{GW} from \ac{ScoX1}, using LIGO detector data from the first
eight months of the fourth observing run of Advanced LIGO, Advanced
Virgo, and KAGRA.  Using the resampling CrossCorr pipeline to search
with a coherence time of $\Tmax=24\un{hr}$, we have set upper limits for
\ac{GW} signal frequencies $25\un{Hz}<f_0<200\un{Hz}$, corresponding
to \ac{NS} spin frequencies of $12.5\un{Hz}<\nu_s<100\un{Hz}$ (or
$15-20\un{Hz}\lesssim\nu_s\lesssim 120-150\un{Hz}$ for $r$-modes).
The sensitivity of the search now surpasses, for the first time the
{\naive} torque balance prediction for plausible signal strengths
regardless of assumptions about the \ac{NS} inclination angle, for
$50\un{Hz}<f_0<200\un{Hz}$.

It is of course quite possible that the spin frequency of \ac{ScoX1}
is above $100\un{Hz}$, with a \ac{GW} frequency above $200\un{Hz}$.
In particular, the distribution of spin frequencies of the observed
\acp{AMXP} appears to be bimodal \citep{Patruno17}, with a ``fast''
population of pulsars, narrowly distributed around
$\nu_s\approx 550\un{Hz}$, which can be explained theoretically by the
presence of GW torques \citep{Gittins19}, and would lead to
gravitational emission at frequencies as high as $1100\un{Hz}$.
Forthcoming analysis of additional data from the fourth observing run
will apply a variety of search methods which balance sensitivity and
robustness to signal model assumptions across a wider range of signal
frequencies.

\section*{Acknowledgments}
\newif\ifcoreonly\coreonlyfalse
\newif\ifkagra\kagratrue
\newif\ifheader\headerfalse
\ifheader
\begin{center}{\bf\Large
\ifkagra
Conference proceedings acknowledgements for \\ the LIGO Scientific Collaboration, the Virgo Collaboration and the KAGRA Collaboration
\else
Conference proceedings acknowledgements for \\ the LIGO Scientific Collaboration and the Virgo Collaboration
\fi
}\end{center}
\fi
This material is based upon work supported by NSF's LIGO Laboratory, which is a
major facility fully funded by the National Science Foundation.
The authors also gratefully acknowledge the support of
the Science and Technology Facilities Council (STFC) of the
United Kingdom, the Max-Planck-Society (MPS), and the State of
Niedersachsen/Germany for support of the construction of Advanced LIGO 
and construction and operation of the GEO\,600 detector. 
Additional support for Advanced LIGO was provided by the Australian Research Council.
The authors gratefully acknowledge the Italian Istituto Nazionale di Fisica Nucleare (INFN),  
the French Centre National de la Recherche Scientifique (CNRS) and
the Netherlands Organization for Scientific Research (NWO)
for the construction and operation of the Virgo detector
and the creation and support  of the EGO consortium. 
The authors also gratefully acknowledge research support from these agencies as well as by 
the Council of Scientific and Industrial Research of India, 
the Department of Science and Technology, India,
the Science \& Engineering Research Board (SERB), India,
the Ministry of Human Resource Development, India,
the Spanish Agencia Estatal de Investigaci\'on (AEI),
the Spanish Ministerio de Ciencia, Innovaci\'on y Universidades,
the European Union NextGenerationEU/PRTR (PRTR-C17.I1),
the ICSC - CentroNazionale di Ricerca in High Performance Computing, Big Data
and Quantum Computing, funded by the European Union NextGenerationEU,
the Comunitat Auton\`oma de les Illes Balears through the Conselleria d'Educaci\'o i Universitats,
the Conselleria d'Innovaci\'o, Universitats, Ci\`encia i Societat Digital de la Generalitat Valenciana and
the CERCA Programme Generalitat de Catalunya, Spain,
the Polish National Agency for Academic Exchange,
the National Science Centre of Poland and the European Union - European Regional
Development Fund;
the Foundation for Polish Science (FNP),
the Polish Ministry of Science and Higher Education,
the Swiss National Science Foundation (SNSF),
the Russian Science Foundation,
the European Commission,
the European Social Funds (ESF),
the European Regional Development Funds (ERDF),
the Royal Society, 
the Scottish Funding Council, 
the Scottish Universities Physics Alliance, 
the Hungarian Scientific Research Fund (OTKA),
the French Lyon Institute of Origins (LIO),
the Belgian Fonds de la Recherche Scientifique (FRS-FNRS), 
Actions de Recherche Concert\'ees (ARC) and
Fonds Wetenschappelijk Onderzoek - Vlaanderen (FWO), Belgium,
the Paris \^{I}le-de-France Region, 
the National Research, Development and Innovation Office of Hungary (NKFIH), 
the National Research Foundation of Korea,
the Natural Sciences and Engineering Research Council of Canada (NSERC),
the Canadian Foundation for Innovation (CFI),
the Brazilian Ministry of Science, Technology, and Innovations,
the International Center for Theoretical Physics South American Institute for Fundamental Research (ICTP-SAIFR), 
the Research Grants Council of Hong Kong,
the National Natural Science Foundation of China (NSFC),
the Israel Science Foundation (ISF),
the US-Israel Binational Science Fund (BSF),
the Leverhulme Trust, 
the Research Corporation,
the National Science and Technology Council (NSTC), Taiwan,
the United States Department of Energy,
and
the Kavli Foundation.
The authors gratefully acknowledge the support of the NSF, STFC, INFN and CNRS for provision of computational resources.

\ifkagra
This work was supported by MEXT,
the JSPS Leading-edge Research Infrastructure Program,
JSPS Grant-in-Aid for Specially Promoted Research 26000005,
JSPS Grant-in-Aid for Scientific Research on Innovative Areas 2402: 24103006,
24103005, and 2905: JP17H06358, JP17H06361 and JP17H06364,
JSPS Core-to-Core Program A.\ Advanced Research Networks,
JSPS Grants-in-Aid for Scientific Research (S) 17H06133 and 20H05639,
JSPS Grant-in-Aid for Transformative Research Areas (A) 20A203: JP20H05854,
the joint research program of the Institute for Cosmic Ray Research,
University of Tokyo,
the National Research Foundation (NRF),
the Computing Infrastructure Project of the Global Science experimental Data hub
Center (GSDC) at KISTI,
the Korea Astronomy and Space Science Institute (KASI),
the Ministry of Science and ICT (MSIT) in Korea,
Academia Sinica (AS),
the AS Grid Center (ASGC) and the National Science and Technology Council (NSTC)
in Taiwan under grants including the Science Vanguard Research Program,
the Advanced Technology Center (ATC) of NAOJ,
and the Mechanical Engineering Center of KEK.
\fi

Additional acknowledgements for support of individual authors may be found in the following document: \\
\texttt{https://dcc.ligo.org/LIGO-M2300033/public}.
For the purpose of open access, the authors have applied a Creative Commons Attribution (CC BY)
license to any Author Accepted Manuscript version arising.
We request that citations to this article use ``A. G. Abac {\it et al.} (LIGO-Virgo-KAGRA Collaboration), \ldots'' or similar phrasing, depending on journal convention.

\software{
\code{LALSuite}~\citep{LALSuite},
\code{numpy}~\citep{harris2020array},
\code{matplotlib}~\citep{matplotlib},
\code{scipy}~\citep{scipy}.
}

This paper has been assigned LIGO Document No.~\dcc.

\appendix

\section{Effects of Elliptical Binary Orbit}

\label{a:eccentricity}

The orbit of \ac{ScoX1} is believed on the basis of modelling of the
binary evolution to be nearly circular ($e\lesssim 10^{-4}$)
\citep{Steeghs2002_ScoX1,Wang2018_PEGS3,Killestein2023_PEGS4}.  Direct
observational constraints on orbital eccentricity are complicated by
systematic effects, and are at the level of $e\le 0.0132$
\citep{Killestein2023_PEGS4}.  Here we consider the possible loss of
\ac{SNR} in the cross-correlation search due to non-zero orbital
eccentricity.

The fractional loss of \ac{SNR} due to searching at parameter values
$\{\lambda_i\}$ for a signal whose true parameter values are
$\{\lambda_i^s\}$ is given in the metric approximation by
\begin{equation}\label{e:mismatch}
  \frac{\ev{\rho}^{\text{ideal}}-\ev{\rho}}{\ev{\rho}^{\text{ideal}}}
  \approx \sum_i \sum_j g_{ij}(\lambda_i - \lambda_i^s)(\lambda_j - \lambda_j^s)
  \ .
\end{equation}
The metric elements for the cross-correlation search are given by
\citep{Whelan2015_ScoX1CrossCorr}:
\begin{equation}
  \label{e:CCmetric}
  g_{ij} = \frac{1}{2}
  \left\langle
    \frac{\partial\Delta\Phi_{KL}}{\partial \lambda_i}
    \frac{\partial\Delta\Phi_{KL}}{\partial \lambda_j}
  \right\rangle_{KL} \ ,
\end{equation}
where $\Delta\Phi_{KL}=\Phi_{K}-\Phi_{L}$ is the \ac{GW} phase
difference for \ac{SFT} pair $KL$ and $\langle\cdot\rangle_{KL}$ is an
average over \ac{SFT} pairs weighted by antenna patterns and detector
noise.

For an \ac{LMXB} in a non-circular orbit, the usual parameters
$\{\lambda_i\}\equiv\{f_0,a\sin i,\Tasc,\Porb\}$ are supplemented by
two additional orbital parameters, the eccentricity $e$ and the
argument of periapse $\omega$ (measured from the ascending node).  For
small values of $e$, it is convenient to work in terms of the
Laplace-Lagrange parameters $\kappa = e\cos(\omega)$ and
$\eta = e\sin(\omega)$ \citep{Lange2001_Timing}.  To lowest order in
$e$, the \ac{GW} phase in \ac{SFT} $K$ [eq.~(4.15) of
\citet{Whelan2015_ScoX1CrossCorr}] becomes
\citep{Messenger2011_Metric,Leaci2015_ScoX1}
\begin{equation}
  \Phi_K = \Phi_0
  + 2\pi f_0\left\{t_K - d_K - a\sin i\left[
      \sin \Psi_K
      + \frac{\kappa}{2}\sin 2\Psi_K-\frac{\eta}{2}\cos2\Psi_K
    \right]
  \right\}
  + \mc{O}(e^2) \ ,
\end{equation}
where $\Psi_K = \frac{2\pi(t_K-\Tasc)}{\Porb}$ is the mean orbital
phase, and $\kappa$ and $\eta$ are both $\mc{O}(e)$.  The \ac{GW} phase
difference $\Delta\Phi_{KL}$ is thus
\begin{equation}\label{e:DeltaPhi}
  \Delta\Phi_{KL} =
  2\pi f_0\left\{\Delta t_{KL} - \Delta d_{KL}
    - a\sin i\left[
      \sin \Psi_K - \sin\Psi_L
      + \frac{\kappa}{2}
      \left(\sin 2\Psi_K - \sin 2\Psi_L\right)
      -\frac{\eta}{2}
      \left(\cos2\Psi_K - \cos 2\Psi_L\right)
    \right]
  \right\}
  +\mc{O}(e^2)
  \ ,
\end{equation}
where $\Delta t_{KL} = t_K - t_L$ is the time difference between
\acp{SFT} $K$ and $L$ and $\Delta d_{KL} = d_K - d_L$ is the
projection onto the line of sight of the separation vector between the
detector positions corresponding to the two \acp{SFT}, and like
$a\sin i$ is measured in light-seconds.

Using the identities
$\sin u - \sin v = 2\cos\frac{u+v}{2}\sin\frac{u-v}{2}$ and
$\cos u - \cos v = -2\sin\frac{u+v}{2}\sin\frac{u-v}{2}$
we can rewrite \eref{e:DeltaPhi} as
\begin{equation}
  \Delta\Phi_{KL} =
  2\pi f_0\left\{\Delta t_{KL} - \Delta d_{KL}
    - a\sin i\left[
      2\cos \overline\Psi_{KL} \sin\frac{\Delta\Psi_{KL}}{2}
      + \left(
        \kappa \cos 2\overline\Psi_{KL}
        + \eta \sin 2\overline\Psi_{KL}
      \right)\sin\Delta\Psi_{KL}
    \right]
  \right\}
  +\mc{O}(e^2) \ ,
\end{equation}
where
$\overline\Psi_{KL}=\frac{2\pi}{\Porb}\left(\frac{t_K+t_L}{2}-\Tasc\right)$
and
$\Delta\Psi_{KL}=\frac{2\pi\Delta t_{KL}}{\Porb}$.

The derivatives with respect to $f_0$, $a\sin i$, $\Tasc$, and $\Porb$
agree to lowest order in $e$ with those given in
\citet{Whelan2015_ScoX1CrossCorr}.  The derivatives with respect to
the Laplace-Lagrange parameters are, to lowest order,
\begin{subequations}
\begin{align}
  \frac{\partial\Delta\Phi_{KL}}{\partial \kappa}
  &= -2\pi f_0 a\sin i \cos 2\overline\Psi_{KL}
    \sin\left(\frac{2\pi}{\Porb}\Delta t_{KL}\right)+ \mc{O}(e) \ ;
  \\
  \frac{\partial\Delta\Phi_{KL}}{\partial \eta}
  &= -2\pi f_0 a\sin i \sin 2\overline\Psi_{KL}
    \sin\left(\frac{2\pi}{\Porb}\Delta t_{KL}\right)+ \mc{O}(e) \ .
\end{align}
\end{subequations}
Arguing, as in \citet{Whelan2015_ScoX1CrossCorr}, that the mean
orbital phase $\overline\Psi_{KL}$ is evenly sampled over the duration
of the run, we can show that the off-diagonal metric elements
\eqref{e:CCmetric}
involving $\kappa$ and/or $\eta$ vanish, and the new diagonal elements
are, to lowest order in $e$,
\begin{equation}
  \label{e:gee}
g_{\kappa\kappa} \approx g_{\eta\eta}
\approx \pi^2 f_0^2 (a\sin i)^2
  \left\langle
    \sin^2\left(\frac{2\pi}{\Porb}\Delta t_{KL}\right)
  \right\rangle_{KL}\ .
\end{equation}
In general, the form of
$\left\langle
  \sin^2\left(\frac{2\pi}{\Porb}\Delta t_{KL}\right)
\right\rangle_{KL}$ depends on the
value of $\Tmax$, scaling as $\Tmax^2$ when $\Tmax\ll\Porb$.  However,
in the present search, $\Tmax$ exceeds $\Porb$, the average approaches
$\frac{1}{2}$, and the metric elements in \eref{e:gee} (like the other
orbital metric elements) saturate, giving us
\begin{equation}
  g_{\kappa\kappa}\approx g_{\eta\eta}\approx \frac{\pi^2 f_0^2 (a\sin i)^2}{2}
  \ .
\end{equation}
Note that this agrees with the ``long-segment regime'' metric of
\citet{Leaci2015_ScoX1} for coherent and semi-coherent $\F$-statistic
based searches.

If we search for a signal with true parameters $\kappa$ and $\eta$
using a template which has $\kappa=0=\eta$, the mismatch is thus
\begin{equation}
  \label{e:eccmismatch}
  \frac{\ev{\rho}^{\text{ideal}}-\ev{\rho}}{\ev{\rho}^{\text{ideal}}}
  = \frac{\pi^2 f_0^2 (a\sin i)^2}{2} (\kappa^2+\eta^2)
  = \frac{\pi^2 f_0^2 (a\sin i)^2 e^2}{2}
  \ .
\end{equation}

\allauthors
\end{document}